\documentclass[12pt]{article}
\usepackage{a4wide}
\usepackage{graphicx}

\usepackage{amstext,amsmath,amssymb,amsfonts,bbm}
\usepackage{epsfig}
\usepackage{hyperref}
\usepackage{amsthm}
\usepackage{color}

\usepackage[all,arc,dvips,arrow,curve,cmactex]{xy}

\usepackage{graphicx}
\usepackage{latexsym}

\newcommand{\be}{\begin{equation}}
\newcommand{\ee}{\end{equation}}
\newcommand{\bea}{\begin{eqnarray}\displaystyle}
\newcommand{\eea}{\end{eqnarray}}

\newcommand{\la}{\langle}
\newcommand{\ra}{\rangle}
\newcommand{\nn}{\nonumber}

\newcommand{\ses}{ {\setminus} } 

\newcommand{\cN}{ {\cal N} } 
\newcommand{\cO}{ { \mathcal{ O } } }
\newcommand{\IC}{{\mathbb C}}

 \newcommand{\cB}{\mathcal{B}}
 \newcommand{\cV}{\mathcal{V}}
 \newcommand{\cE}{\mathcal{E}}

\newcommand{\cZ}{ \mathcal{Z} }

\newcommand{\diag}{ {\rm Diag} } 
\newcommand{\Sym}{ {\rm Sym} } 
\newcommand{\plog}{ {\rm Plog} }

\newcommand{\ncol}{{\rm noncolor}}
\newcommand{\sym}{{\rm sym}}
\newcommand{\col}{{\rm{sc}}}
\newcommand{\conex}{{\rm connected}}
\newcommand{\mC}{\mathbb{C}} 
\newcommand{\Tr}{{\rm Tr}}
\newcommand{\tr}{{\rm tr}}
\def\s{ \sigma }

\newtheorem{definition}{Definition}

\newtheorem{proposition}{Proposition}

\begin{document}

\begin{flushright}
QMUL-PH-13-08\\
ICMPA-MPA/2013/008
\end{flushright}

\bigskip

\begin{center}

{\Large\bf Counting Tensor Model Observables }
 \medskip 

{ \Large \bf  and Branched Covers of the 2-Sphere  }

\bigskip

{
Joseph Ben Geloun$^{a,c,*}$
 and Sanjaye Ramgoolam$^{b , \dag}  $}

\bigskip

$^a${\em Perimeter Institute for Theoretical Physics, 31 Caroline
St N} \\
{\em ON N2L 2Y5, Waterloo, ON, Canada} \\
\medskip
$^{b}${\em School of Physics and Astronomy }\\
{\em Queen Mary, University of London }\\
\medskip
$^{c}${\em International Chair in Mathematical Physics
and Applications}\\
{\em ICMPA--UNESCO Chair, 072 B.P. 50  Cotonou, Republic of Benin} \\
\medskip
E-mails:  $^{*}$jbengeloun@perimeterinstitute.ca,
\quad $^{\dag}$ s.ramgoolam@qmul.ac.uk

\begin{abstract}

Lattice gauge theories of permutation groups with a simple topological action (henceforth 
permutation-TFTs)  have recently found 
several applications in the combinatorics of quantum field theories (QFTs). 
They have been used to solve  counting problems of 
Feynman graphs in QFTs and ribbon graphs of large $N$,  often revealing 
inter-relations between different counting problems.  In another  recent development, 
tensor theories   generalizing   matrix theories  have been actively developed as models of 
random geometry in three or more  dimensions. Here, we apply  permutation-TFT methods to count
gauge invariants for  tensor models (colored as well as non-colored),  
exhibiting a relationship with counting problems of 
branched covers of the  2-sphere, where the rank  $d$ of the tensor gets related to a number of 
branch points.  We give explicit generating functions for the relevant counting and 
describe algorithms for the enumeration of the invariants. As well as 
the classic count of Hurwitz equivalence classes of branched covers with fixed branch points, 
collecting these under  an equivalence of permuting the   branch points is relevant to the color-symmetrized  tensor invariant counting. 
We also apply the permutation-TFT methods to obtain some formulae 
for correlators of the tensor model invariants.

\end{abstract}

\end{center}

\noindent  Key words: Matrix/tensor models,
tensor invariants, topological field theory, branched covers, 
graph enumeration, permutation groups.

\newpage 

\tableofcontents

\section{Introduction}

Motivated by the problem of understanding the precise dictionary between observables in string theory and in 
gauge theory, in the context of gauge-string duality \cite{'tHooft:1973jz,malda}, permutation group techniques have recently 
been used to solve a variety of problems in the combinatorics of single- and multi-matrix models
 \cite{cjr,BBFH, KR1,KR2,BHR1,BHR2, BCD1,BCD2,quivcal,doubcos}.  
 
 While the matrix models often  arise from the study  of particular sectors of four dimensional 
  $ \cN=4$  super-Yang-Mills theory, or other four dimensional gauge theories, it has been fruitful  to revisit, with these permutation techniques, the study of matrix models as mathematical 
  models of gauge-string duality  in their own right \cite{Gal,Gop1,Gop2,Garner:2013qda}. This line of research 
   draws key ideas from the discovery that the
 large $N$ expansion of two dimensional Yang-Mills (YM) theory, can be reformulated as a string theory, a link where permutations 
 play a crucial role \cite{GRT1,GRT2,CMR1,CMR2,dadprov}. The   exact partition function of $U(N)$ 2dYM on a Riemann surface $ \Sigma_g$  \cite{Migdal:1975zg} 
can be expanded in $1/N$ and the coefficients in the expansion were recognized as counting 
  holomorphic maps between Riemann surfaces $ \Sigma_h \rightarrow \Sigma_g$.  The power of $N$ is related to the genus of the covering space $ \Sigma_h$, 
which is interpreted as the string worldsheet.

  There are three main elements to this YM-string connection. 
The first is    the mathematical fact of { \it Schur-Weyl duality}  which relates the world of unitary groups,
 more generally classical groups, to the world of permutations. 
 The second is   { \it two dimensional topological field theory of permutations} (permutation-TFT),  
 a simple physical construction based on lattice gauge theory, 
 with symmetric groups as gauge groups, where edges variables take values in a symmetric group. 
  The plaquette weight of the lattice theory is  a simple delta function, which gives one when the edge variables around 
  the plaquette multiply to one, and gives zero otherwise.     The third is the link between
   {\it permutations and covering spaces}, a basic fact of
 algebraic topology. Now in two dimensions, branched covers are equivalently holomorphic maps, leading to 
 deep links between combinatorics and  complex geometry in  the form of the Riemann existence theorem. Combining these ingredients 
 leads to an interpretation of the permutation sums that appear in the large $N$ expansion of 2dYM in terms of 
 spaces of branched covers, equivalently  holomorphic maps, called Hurwitz spaces. 

  The link between  permutations and strings - at a topological level - 
   is of course rather simple, and deep in this simplicity :  strings winding around a circle 
  have a winding number. For a  fixed total winding number $n$, multi-string configurations 
  contain a configuration for every partition of $n$.  This has motivated the investigation of Feynman graph counting problems in QFT 
 in terms of permutations,  including situations without large $N$ \cite{FeynCount}. 
 The structure of a graph can be coded using numbers to give labeled structures,  in such a way that there is an action of 
 permutation groups (of re-arrangements of the numbers) on  the labeled structures, 
 and the counting of the graphs  involves modding out by certain permutation equivalences. 
 This leads to the combinatoric description of graphs in terms of double cosets, which was heavily exploited in \cite{FeynCount}. 
Problems of refined graph counting, in this case of graphs embedded in Riemann surfaces, were studied 
in \cite{refcount} using further techniques such as graph quotients. 
The central role of permutation-TFTs continues to persist in these cases. As a unifying 
description of diverse counting problems, the permutation-TFTs often reveal surprising connections, a notable one being 
the link between the counting of vacuum graphs in quantum electrodynamics and that of ribbon graphs, which are normally  encountered 
in a large $N$ context. The cyclic orientation provided by the electron circulating  in loops, 
can be mapped to a problem of  graphs with vertices equipped with cyclic orientation,  which are precisely ribbon graphs. 
While this is a  good way to understand the surprising  link in retrospect, it is easiest
 to derive it  by manipulating some delta functions over symmetric groups.

\begin{figure}[ht]
\begin{center}
 \resizebox{!}{4cm}{\includegraphics{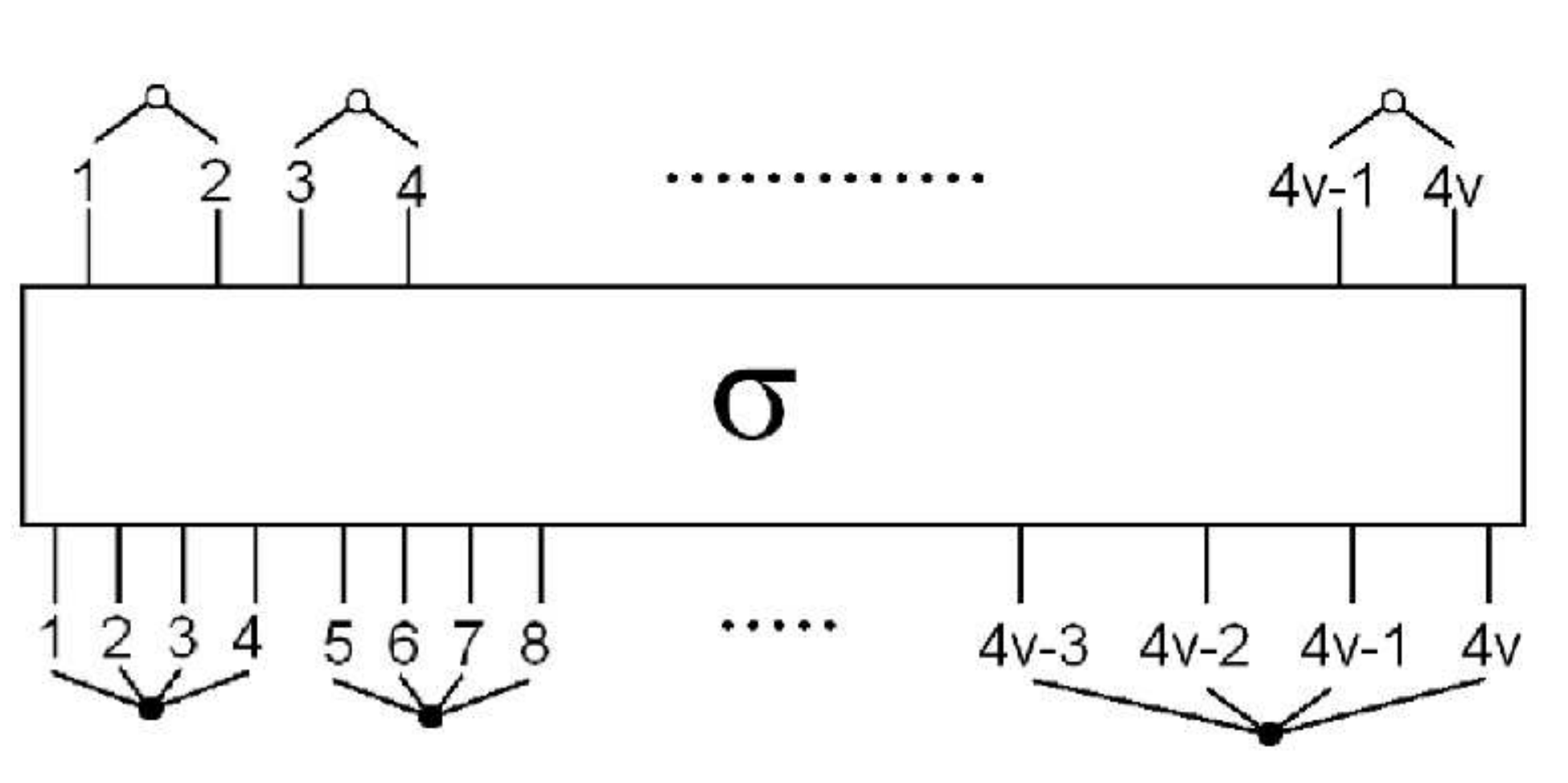}}
\caption{Double coset connection}
 \label{fig:dcoset}
\end{center}
\end{figure}

In this paper, we will undertake some counting problems motivated by tensor models, 
using the framework of permutation-TFTs, and we will find that this framework  continues to be a source
of non-trivial links between apparently very different counting problems. 
Let us review a little more explicitly some concepts  from  \cite{FeynCount}, which will set the stage 
for our current investigations.  A Feynman graph can be coded in terms of labeled combinatoric data, by first
introducing in the middle of all the existing edges a new type of vertex to get a new graph. 
We can call the formerly existing 
vertices - black vertices,  and the newly introduced bivalent  vertices -  white vertices. 
Now label the edges of the new graph with integers $ \{ 1, 2, \cdots , 2 d \} $, where 
$d$  is the number of edges of the original graph. Next, cut along all these 2d edges. 
All graphs with a fixed vertex structure can be obtained by re-connecting these cuts. 
The different reconnections can be parametrized by a permutation $ \sigma \in S_{2d}$. 
This is illustrated in Figure \ref{fig:dcoset} for the case  where we have $v$  4-valent vertices 
in the original graph and $ d= 4v$.  Different permutations can give the same graph if they are related by 
equations of the form $ \sigma' = \gamma_1 \sigma \gamma_2$. The $\gamma_1 , \gamma_2$ 
live in subgroups $H_1, H_2$  of $S_{2d} $ related to the symmetries of the black vertices and of the
white vertices respectively. This allows us to count Feynman graphs by counting points in double cosets of 
permutation groups. 

Burnside's lemma leads to formulae for counting  these equivalence classes
as sums of some delta functions over symmetric groups. 
These can in turn be recognized as the partition functions of
permutation-TFT on a cylinder, with $S_{2d}$ gauge group,  and with boundary observables related to
$H_1 , H_2$. The  above framework for relating graph counting to permutation groups and 
in particular  permutation-TFTs is rather general.

  Graph counting has also come into the centre of attention from a completely different perspective, namely random tensor models. 
 Graphs are related (see for example   \cite{BGR1202})  to the counting of tensor invariants  
 a problem with classical origins   \cite{mukai,little}. 
 Tensor models have been proposed as a way to understand higher dimensional random 
geometry \cite{ambj3dqg,mmgravity,sasa1,Oriti:2006se}, generalizing the powerful results connecting matrix models to two dimensional 
quantum gravity from the eighties/nineties \cite{Di Francesco:1993nw}. 
 A tensor model is defined via a field
which is a rank $d$ tensor over an abstract multi-dimensional representation space. In a dual space, a rank $d$ tensor is viewed as a
$(d-1)$-simplex. The interaction in such models is dually described by a 
$d$-simplex and is formed by the gluing of $(d-1)$-basic simplices 
along their $(d-2)$-boundary simplex. For example, 
if $d=2$, the field can be a real matrix $M$ representing a 1-simplex or 
a segment; the simplest interaction is of the form of an invariant
$\Tr[ M^3]$ and represent a triangle formed by the gluing of
1-simplices along their 0-simplex boundaries. This is the simplest
non trivial matrix model. The simplest higher rank extension of this
model, is a rank 3 tensor model. Here, the field is a rank 3 tensor 
representing a 2-simplex or triangle. The interaction is obtained 
by a specific contraction of tensor fields and represents a $3$-simplex
or tetrahedron formed by the gluing of triangles along 
their 1-simplex or boundary segments. Generally, in a rank $d$ model, 
a Feynman graph corresponds to a simplicial complex obtained
from the gluing of $d$-simplexes along their $(d-1)$-boundary.

Recent work has focused 
on colored tensor models \cite{color,Gurau:2011xp} where the  $1/N$ expansion has been developed \cite{Gur3,GurRiv,Gur4}.
This has triggered a plethora of new results on higher dimensional statistical mechanics and renormalizability of tensor models 
 \cite{Gurau:2011kk,Gurau:2011tj,Gurau:2012vk, Gurau:2012ix,Bonzom:2011zz, Bonzom:2011ev,Gurau:2013cbh, Gurau:2013pca, 
BenGeloun:2011rc,Geloun:2012fq,Geloun:2013saa}.  ``Melonic graphs,'' which can be counted 
 by  mapping to a tree counting problem, with counting functions given by generalized Catalan numbers 
 have played a central role, notably in connection with solving Schwinger-Dyson equations \cite{Bonzom:2011zz}.
Specific types of melonic tensor invariants have been used in QFT to determine renormalizable actions 
 \cite{BenGeloun:2011rc,Geloun:2013saa}.

In this paper, we will consider tensor models where the basic fields are $\Phi_{ i_1, \cdots , i_d } $
and $ \bar \Phi_{ j_1 , \cdots j_d } $. The indices  $i_1 , \cdots i_d $ transforming as $ \otimes_{a=1}^n  V_a  $  of 
$U(N_a)^{ \times d } $ , while $ j_1 \cdots j_d $ transform as $ \bar V_a^{ \otimes n } $, with $V_a$ being the fundamental 
of $U(N_a)$ and $\bar V_a$ the anti-fundamental. The emphasis will be on the complete enumeration 
of tensor invariants, given specified gauge invariance constraints. We will focus our attention of 
the case $d=3,4$, and take $n$ to be the number of $\Phi$'s, which has to be equal to the number of 
 $\bar \Phi$'s. Based on the expectations from \cite{FeynCount,refcount} we find that these counting problems 
 can be expressed  neatly in terms of permutation-TFTs. And we find that these problems, for any $d$,  can be mapped to 
 the counting of branched covers of the two sphere. The parameter $d$ appears as the number of branch points 
 on the 2-sphere. 

These formulations in terms of TFTs and branched covers allow the expression of 
the counting in terms of extracting coefficients of certain multi-variable generating functions. 
 These expressions can be evaluated to 
high orders with the help of Mathematica, where the enumeration of the tensor invariants by hand becomes hopeless. 
Another useful piece of software is GAP \cite{GAPpage}, which gives not only the numbers of invariants, but 
can also store the detailed information about the structure of the invariant in the form of 
some permutation data, once the correct permutation formulation of the tensor counting 
problem has been found.

The plan of the paper is as follows. The next section reviews the definition of unitary tensor invariants (sometimes referred to, 
 simply as tensor invariants or even trace invariants) 
and tensor models. 
Section \ref{sect:Counting} deals with the counting of invariants that can be built from 
$d$-index tensors $ \Phi_{ i_1 , \cdots , i_d } , \bar \Phi_{ j_1 , \cdots , j_d } $, when we have 
$n$ copies of $\Phi $ and $\bar \Phi$. We first formulate the problem in terms of counting 
invariants of an action of $U(N)^{ \times d } $ on a certain symmetrized tensor product of
fundamental representations. This is  mapped to  a counting of a $d$-tuple of  permutations subject to 
certain constraints, which are themselves given by the action of two permutations. 
These two permutations correspond to the symmetries of re-ordering the $ \Phi$'s and $\bar \Phi$'s respectively. 
This problem is expressed in terms of sums over delta functions over symmetric groups, which are then simplified 
to yield a problem of counting a sequence of just $d-1$ permutations, subject to an equivalence given by one permutation. 
This leads to a solution 
of the counting in the form of sums over partitions, weighted by powers of the symmetry factors of 
the partitions (see Equation \eqref{ncopdtens}).  We distinguish connected and disconnected invariant counting, which are related 
by the plethystic Log function.  Section \ref{sect:connect} interprets the symmetric group delta functions arising 
in the solution above in terms of topological lattice gauge theory on a certain complex. The simplification 
 is shown to be related to a  coarsening of the complex, which leaves the answer invariant because 
of the topological invariance of the lattice theory. The final permutation problem involving $d-1$  permutations 
with one conjugation  constraint is explained, using the classic Riemann existence theorem, to be related to the counting 
of branched covers of the sphere, equivalently to holomorphic maps from Riemann surfaces to the sphere. 

Section \ref{sect:repack} describes a  {\it  color symmetrized }   version 
of the counting problem of Section \ref{sect:Counting}. This is based on the fact that the 
counting of colored tensor invariants for rank $d$  admits an $S_d$  permutation 
symmetry of renaming the colors, so it is natural 
to count equivalence classes under this symmetry. This problem is expressed in precise form in terms of 
$U(N)^{\times d }$ invariants in an appropriate vector space. Again, since $U(N)$  invariants are generated 
by products of $\delta_{ij}$, we can count them by parametrizing the possible ways the $i$'s go with the $j$'s, 
which is given by permutations $\sigma_a$  (for $ a $ going from $1$ to $d$). The color-symmetrized counting 
involves imposing a further equivalence of permuting the $ \sigma_a$. We solve this using the permutation group algebra techniques, the upshot being  simple elegant formulae in 
terms of delta functions over symmetric groups, leading to generating functions involving multiple variables. 
There is a subtlety in the relation between connected and disconnected case, so that the connected counting is no 
longer given by taking a plethystic log. This subtlety is explained. 

 Section \ref{sect:other} turns to the counting of general tensor invariants where contraction between the different 
 $i$-indices on $\Phi$ can occur with $j$ indices on $ \bar \Phi$, irrespective of the positions of these indices. 
 This distinguishes the counting from the ``colored-case''  where the different slots along the $\Phi$ 
 and $ \bar \Phi$  are distinguished, so we may call this non-colored counting. This is a problem in invariants of 
 $U(N)$ acting on an appropriate vector space, rather than $U(N)^{\times d } $. A variation where the 
 tensors $ \Phi , \bar \Phi $ are symmetric is also solved. Section \ref{sec:correlators} gives some formulae 
 for correlators  related to the counting of Section \ref{sect:Counting}. 
 
Section \ref{sec:discussions} gives a summary of our results and avenues for future research. The discussion includes, as well, some observations
 on the relations between color-symmetrized counting of invariants and the counting of braid orbits  of branched covers, 
a subject that is studied from completely different motivations by pure group theorists. 
Appendix \ref{app-group-basics}  gives a short review on group actions, including Burnside's lemma and some key facts about 
the symmetric group.  Appendix \ref{app:s2n} proves some formulae stated  in the main text.  Appendix \ref{app:corrs} provides details about derivations of formulae 
for  correlators in Gaussian tensor models given in Section \ref{sec:correlators}. 
Appendix \ref{app:gap}  contains some GAP and Mathematica  codes used to obtain the explicit counting sequences\footnote{These can certainly be improved in efficiency but are included for illustrations.}. 
Some of these are identified with known ones in OEIS \cite{oeis} others are not in OEIS.

\section{Tensor model invariants: a review}
\label{sect:TIM}

In this section, we review the  construction of unitary tensor
invariants and their graphical representation. 
The results presented here are largely based on \cite{Gurau:2012ix}.  We also discuss the simplest way 
to introduce tensor models and their Feynman graphs.

\subsection{Tensor invariants}
\label{subsect:ti}

Let $V_1$, $V_2$, \dots, $V_d$ be some complex vector spaces of dimensions $N_1$, $N_2$, \dots $N_d$. 
Consider rank $d\geq 2$ covariant tensors $\Phi$ 
with components $\Phi_{i_1 , \cdots , i_d  }$ transforming as  $\otimes_{a=1}^d V_a$
with $i_a  \in \{1,\dots, N_a\}$, $a=1,2,\dots, d$, with no symmetry assumed under permutation of their indices. 
These tensors transform under the action of the  tensor product of fundamental representations of unitary groups $\otimes_{a=1}^{d} U(N_a)$
where each unitary group $U(N_a)$ acts on a tensor index $i_a $
independently. The complex conjugate of $\Phi_{ i_1 i_2  \dots i_d }$
is a contravariant tensor of the same rank and is given by $\bar{\Phi }_{i_1 i_2 \dots i_d}$. We have the following transformation:
\bea
\Phi_{i_1i_2 \dots i_d} &=& \sum_{j_1, \dots, j_d} 
U^{(1)}_{i_1 j_1} U^{(2)}_{i_2 j_2} \dots U^{(d)}_{i_d j_d}  \Phi_{j_1 j_2 \dots j_d } \cr
\bar{\Phi}_{i_1i_2 \dots i_d} &=& \sum_{j_1, \dots, j_d} 
\bar{U}^{(1)}_{i_1 j_1} \bar{U}^{(2)}_{i_2j_2} \dots \bar{U}^{(d)}_{i_dj_d}
  \bar{\Phi }_{j_1j_2 \dots j_d } 
\label{tut}
\eea
where $U^{(a)} \in U(N_a)$ and  may be very well all distinct. 
In the next discussion, we will be primarily  interested in $d\geq 3$.  

Invariants with respect to the unitary action 
\eqref{tut} built on tensors can be obtained  by contracting, 
in all possible ways, pairs of covariant and contravariants tensors. 
It turns out that these contractions are in bijection with closed $d$--colored graphs
that we must  now introduce. 

A bi-partite closed $d$-colored graph is a graph $\cB=(\cV(\cB),\cE(\cB))$ 
that is a collection $\cV(\cB)$ of vertices with fixed valence (or degree or coordination) $d$ and set $\cE(\cB)$ of edges, with incidence relation between 
edges and vertices, such that 

- $\cV(\cB)$ can be partitioned into two disjoint  sets $\cV^+$ and $\cV^-$, of equal size, 
such that each edge $e$ is may only connect a vertex $v^+\in \cV^+$
and a vertex $v^- \in \cV^-$ (this is the bi-partite property); 

- the graph has a $d$-line coloring $\alpha$, that is an assignment of a color to each edge, $\alpha: \cE(\cB)\to \{1,2,\dots,d\}$, 
such that two adjacent edges cannot have the same color (two edges are called adjacent
if they are incident to a same vertex). Note that $\alpha^{-1}(i)$
is the subset of lines of color $i$. 

The fact that the graph is closed simply implies that 
the number of edges in the graph fully saturates the 
valence of the vertices: $2\cE(\cB) = d\cV(\cB)$.

\begin{figure}[h]\begin{center}
     \begin{minipage}[t]{.8\textwidth}
\includegraphics[angle=0, width=12cm, height=3.5cm]{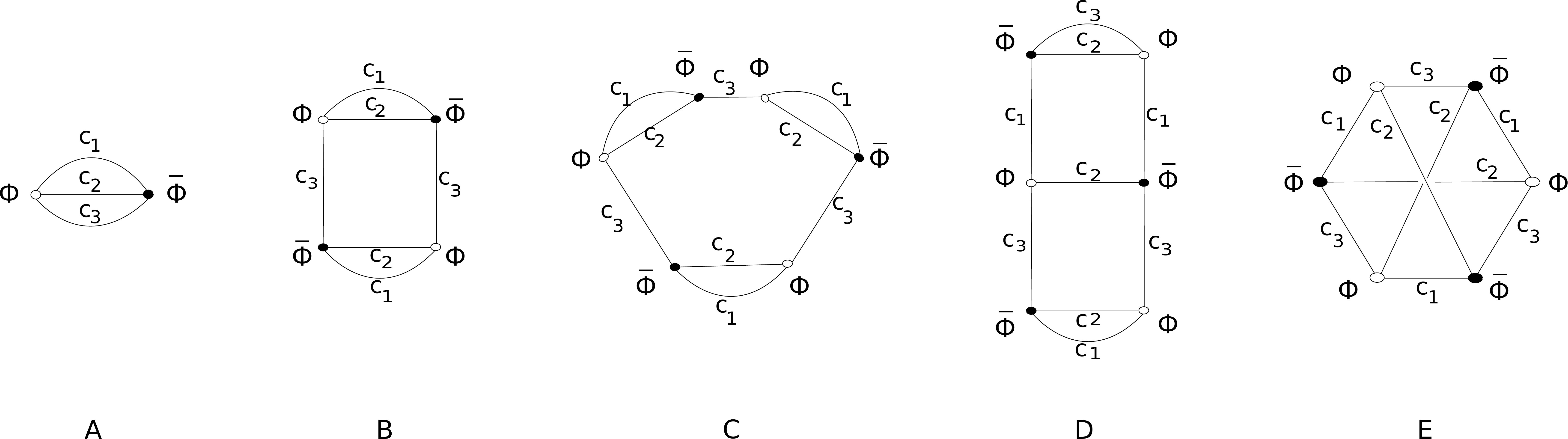}
\vspace{0.3cm}
\caption{ {\small Some rank $d=3$ tensor invariants}} \label{fig:tensinv}
\end{minipage}
\end{center}
\end{figure}

One can construct the graph associated with a tensor 
invariant built from the contraction
of some tensors in the following way. Consider $\Phi_{i_1 \dots i_d}$ 
(respectively, $\bar{\Phi}_{i_1\dots i_d}$)
and assign it to a vertex $v^+ \in \cV^+$
(respectively, to a vertex $v^- \in \cV^-$). The position of 
an index in the tensor becomes a color: $i_a$ has the color $a$. 
The contraction of an index $i_a$ of some $\Phi_{\dots i_a \dots}$ with and index $j_a $
of some $\bar \Phi_{\dots j'_a \dots}$ is represented by a line of color $a$ between
a vertex $v^+$  associated with $\Phi$ and a vertex $v^-$ associated
with $\bar{\Phi }$. Some examples are provided in Figure \ref{fig:tensinv}. 
The trace invariant associated with $\cB$ is given by 
\be\label{traceinv}
\Tr _{\cB} (\Phi  ,\bar \Phi )  = \sum_{ i , j } \delta^{\cB}_{i, j } \prod_{v,v' \in \cV(\cB)}
\Phi_{i_v} \bar{\Phi}_{j_v} \,, \qquad \quad  
\delta^{\cB}_{i,j } = \prod_{a=1}^d \; \prod_{l^a \in \alpha^{-1}(a)} 
\delta_{i^a_{v^+(l^a)} , n'^{a}_{v^-(l^a)}}
\ee
where in the formula, the sum is performed over all indices 
of the tensors, the function $\delta^{\cB}_{i,j}$ implements the
$d$-line coloring or contraction between tensor indices, such that, given a line
$l^a$ of color $a$ incident to vertices $v^+(l^a)$ and $v^-(l^a)$, 
the indices $i^a_{v^+(l^a)}$ must be equal to $j^{a}_{v^-(l^a)}$.  
One can check the formal expression
\be
\Tr _{\cB} (\Phi^{U} ,\bar \Phi^{\bar{U}}) 
=\Tr _{\cB} (\Phi ,\bar \Phi )
\ee
with $\Phi^{U}$ stands for the transformed of $\Phi $ with respect to 
the unitary action \eqref{tut}. 
The trace invariant may factorize over the connected components
of $\cB$. For instance in Figure \ref{fig:tensinv}, combining
graphs A and B generates a new rank 3 disconnected invariant
made with six tensors.  

Finally, we must emphasize that colored graphs of this kind 
are dual to $d$-dimensional abstract simplicial pseudo-manifolds \cite{color}.
Such a feature is important in the framework of tensor models.   
In the same way that the study of matrix models provides the statistical sum of random triangulations of Riemannian surfaces 
and turned out to be important to solve 2D quantum gravity, tensor models generate random triangulations of higher dimensional objects and 
address gravity in dimension higher than 2. The colored tensor model introduced in  \cite{color} yields a first step towards 
a clearer understanding of the type of ``regular'' triangulations that can be generated by the partition function using colored tensors. The next section formally introduces
the generic type of tensor models.

\subsection{Tensor models}
\label{subsect:tm}

The simplest form of rank $d$ tensor models are described by an action with complex tensor field $\Phi_{i_1 \dots i_d}$ with kinetic term 
\be
S^{\rm kin} =   \sum_{\{ i_a \} } \bar\Phi_{ i_1  \dots i_d }\,\Phi_{ i_1   \dots i_d } 
\ee
In the specific instance $S^{\rm kin}$ corresponds to a mass term. 
Certainly, more elaborate kinetic terms can be constructed.

A typical $(\bar\Phi\Phi)^p$ interaction in such a model may be 
written as
\be 
S^{ \rm inter  } = \lambda_V  \sum_{ \{ i^{(p)}_a , j^{(p)}_a \} }
V ( \{ i^{(p)}_a , j^{(p)}_a \} ) \prod_{p=1}^n \bar \Phi_{ i^{(p)}_1 \cdots i^{(p)}_d  }
  \Phi_{ j^{(p)}_1 \cdots j^{(p)}_d  } 
\ee
where  $\lambda$ is a coupling constant and $V$  is constructed from Kronecker delta's 
and determines the precise form of the interaction. In 1-matrix theory, interaction terms 
are, say at order $3$,  of the form $\tr (M^3), [\tr (M^2)](\tr M), (\tr M)^3$: at order $n$ 
there are $p(n)$ possible interaction terms (number of partitions of $n$).
 The enumeration of tensor invariants we give in subsequent 
sections allows a group theoretic characterization  of the interaction terms at each order 
for tensor models and gives a  number $Z_d(n)$ which replaces $p(n)$ when we go from matrix models to tensor models.  
  Particular forms of $V$ might lead to models with different properties. 
For instance, discussing perturbative renormalizability, the type of contractions implemented by $V$ 
should be of form of trace invariants of the melonic kind \cite{Geloun:2013saa}. 

The partition function associated with the type of tensor
model can be written 
\be
Z= \int d\Phi d \bar \Phi e^{- S^{\rm kin} -  S^{\rm inter}(\bar\Phi, \Phi)}
\label{parz}
\ee
Either at the Gaussian limit $\lambda=0$ or, in the perturbative picture, by perturbing around the Gaussian measure,  
the several types of countings that we will discuss in the following
are useful for the understanding of the $2P$ correlation function
issued from tensor models as
\be
\la \bar\Phi_{I_1}\Phi_{I'_1}\dots 
\bar\Phi_{I_P}\Phi_{I'_P} \ra = \int   d\Phi d \bar \Phi    ~~~     \bar\Phi_{I_1}\Phi_{I'_1}\dots 
\bar\Phi_{I_P}\Phi_{I'_P}  \; e^{ -  S^{\rm kin} - S^{\rm inter}(\bar\Phi, \Phi)}
\ee
where $I_i$ are multi-indices. The external data $(I_1,I_2,\dots, I_P)$
and $(I'_1,I'_2,\dots, I'_P)$
are associated with external boundary topological data of the 
simplex corresponding to the collection of fields $(\bar\Phi_{I_1},\Phi_{I'_1},\dots, \bar\Phi_{I_P}, \Phi_{I'_P})$. Referring to the renormalizable tensor models, it has been proved that,
for a primitively divergent correlation function, these momentum data should match with melonic tensor invariant contractions of the same form of the vertex in the initial action.  Once again, unitary invariants 
play a central role in this context.

\section{Counting invariants in colored tensor models } 
\label{sect:Counting}

For simplicity, we start the discussion by the rank $d=3$ case, 
the general situation $d \geq 3$ can be easily inferred from this 
point. 

\medskip 

\subsection{ Tensor invariants as $U(N)^d $ group action invariants} 

 Consider a colored tensor $\Phi$ of rank $d$, where the 
 indices are colored. Making the indices explicit, we have 
$ \Phi_{ i_{1} \cdots i_d  } $.
We want to know the number $Z_3(n)$ of invariants in the form 
\eqref{traceinv} that one can build from 
$ n$ copies of $ \Phi$ and $n$ copies of $ \bar \Phi$. 

This can be formulated as a problem in invariant theory.
Given a $U(N)$ representation $V$, there is a one-dimensional 
space of  linear maps from $ V \otimes \bar V $ to $\mC$ such that 
\bea 
\delta\, | e_i > \otimes | \bar e_j > = \delta_{ ij } 
\eea
which are invariant in the sense that  
\bea 
 \delta ( U \otimes U )  = \delta \,,\qquad  \forall U \in U(N) 
\eea
This follows since 
\bea 
 \delta ( U \otimes U )\, | e_i > \otimes | \bar e_j > &=& U_{k i }   ( U^{\ast } )_{lj} \,
  \delta ( | e_k > \otimes |\bar e_l   > ) \cr 
&=&  U_{ki}  ( U^{\ast} )_{kj }  = \delta_{ij}  
\eea
Given $\otimes_{a=1}^d V_a $ which is a representation of $ U(N)^{\times d}$ of dimension $N^d$
 (note that we could equally well work with $U(N_1) \times U(N_2) \times \cdots \times  U(N_d)$,
  in which case we have dimension $N_1 N_2 \cdots  N_d  $), consider 
\bea 
&& W = \Sym ( V_1 \otimes V_2 \otimes \cdots\otimes  V_d  )^{ \otimes n }  \cr 
&& \bar W =  \Sym ( \bar V_1 \otimes \bar V_2 \otimes \cdots\otimes \bar V_d  )^{ \otimes n }
\eea
The $\Sym$ indicates that we are symmetrizing the $n$ copies
(in other words these define indistinguishable copies) 
The first counting problem we solve is to find the dimension of the space of 
invariants in $ W \otimes \bar W $. We will assume $N > n $, otherwise  there are finite $N$ corrections
which we leave for future investigation (see more comments on this in the discussion section). 

Now $W$ has an action of $S_d$ of permuting  $V_1 \otimes V_2 \otimes  \cdots \otimes V_d $, likewise for $ \bar W$. 
We can  define a linear operator for each $ \alpha \in S_d$, denoted $ \rho_W ( \alpha  ) $ 
acting on $W$ and a linear operator $ \rho_{ \bar W } ( \alpha )$ acting on $ \bar W $. 
Consider the  $S_d$-symmetrizer acting on $ W \otimes \bar W $
given by 
\be\label{s3sym}
{ 1 \over d! } \sum_{ \alpha \in S_d }  \rho_W ( \alpha )  \otimes \rho_{ \bar W }  ( \alpha )  
\ee
The second problem in invariant theory is to count the dimension of the space of $ U(N)^{\times d}$ invariants in the image of the above symmetrizer. 
This is the color-symmetrized counting we address in Section \ref{sect:repack}.  

\subsection{ Tensor invariants for $d=3$ and permutation double coset } 

At this point we will, for concreteness,  specialize the discussion 
to $d=3$, although it will be clear how the steps generalize to general $d$. 
Returning to the first problem, the invariants are generated by the different ways of contracting 
the different copies of $V_a$  in $W$ with the copies of $\bar V_a$ in $\bar W $.  Diagrammatically, one may think about all the possible contractions 
between $n$ tensors simply as the possible parings in the way given in Figure \ref{fig:sss0}. 
\begin{figure}[h]\begin{center}
     \begin{minipage}[t]{.8\textwidth}\centering
\includegraphics[angle=0, width=8cm, height=3cm]{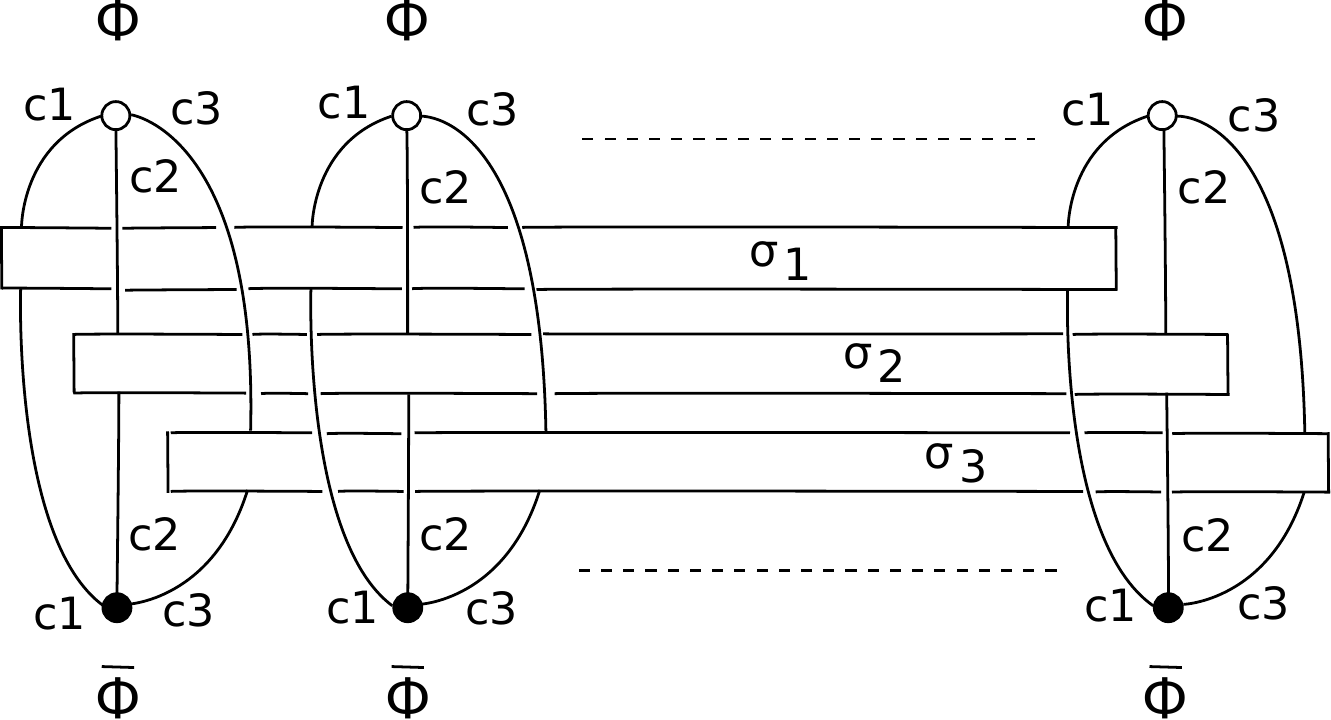}
\vspace{0.3cm}
\caption{ {\small Diagrammatic tensor contraction defining $(\s_1,\s_2,\s_3)$}} \label{fig:sss0}
\end{minipage}
\end{center}
\end{figure}
In other words, the determination of possible 
graph amounts to count triples 
\be 
( \s_1 , \s_2 , \s_3 ) \in  ( S_n \times S_n \times S_n ) 
\ee
with equivalence 
\be
\label{gamma-equivs} 
( \s_1 , \s_2 , \s_3 ) \sim ( \gamma_1 \s_1 \gamma_2 , \gamma_1 \s_2 \gamma_2 , \gamma_1 \s_3 \gamma_2 ) 
\ee
where $ \gamma_i \in S_n$. 
 Thus, we are counting points in the double coset 
\be
 \diag ( S_n ) \ses  ( S_n \times S_n \times S_n ) / \diag ( S_n ) 
\ee
We denote the number of point in this double coset as $ Z_3(n)$. 
For general subgroups $ H_1 \subset G , H_2 \subset G $, the cardinality  of this double coset is given by 
\be 
|H_1\ses G / H_2| = 
{ 1 \over |H_1|  | H_2 |   } \sum_{  C } Z_C^{ H_1 \rightarrow G } Z^{ H_2 \rightarrow G }_C \,\Sym ( C )  
\label{hgh2}
\ee
The sum is over conjugacy classes of $G$, and $Z_C^{ H \rightarrow G }$ is the number of elements of $H$ in the conjugacy class $C$  of $G$.  
This formula appears in the context of graph counting in \cite{Read} 
and is used for a variety of Feynman graph problems in  \cite{FeynCount}.

Let us explain the proof of this formula using Burnside's lemma reviewed in Appendix \ref{app-group-basics}.
 We think 
of the double coset as the number of orbits of the $H_1 \times H_2 $ action 
on $G$. The fixed-point counting formula for the number of orbits becomes 
\bea\label{doubcosdelta}  
|H_1\ses G / H_2| = { 1 \over |H_1| |H_2| } 
\sum_{ h_1 \in H_1 } \sum_{ h_2 \in H_2 } \sum_{ g \in G } 
\delta   ( h_1 g  h_2 g^{-1} ) 
\eea
where $\delta $ is the delta function over the group $G$, equal to $1$ if its argument 
is the identity element and $0$ otherwise.  
This means that $h_1 , h_2 $ have to be in the same conjugacy class of $G$.
Now organize the sums according to conjugacy classes $C$ of $G$. 
The number of elements in conjugacy class $C$ 
from $H_1$  and $H_2$ are denoted $ Z^{H_1}_C , Z^{H_2}_C $.
So the counting has a factor $   Z^{H_1}_C  Z^{H_2}_C$ 
from the $ h_1, h_2$ sums.
For each such pair, there are $ \Sym (C) $ possible $ g $'s. 
Hence we get the above formula.

The conjugacy classes of $S_n \times S_n \times S_n $ are entirely determined by a triple $(p_1 , p_2 , p_3 )$ where each $p_i$ is a partition
  of $n$ (see Proposition \ref{prop:cc} in Appendix \ref{app-group-basics}).
 This correspondence holds because each conjugacy class is determined by a cycle structure. 
Now, the diagonal subgroup produces conjugacy classes
$( p , p , p )$. So applying \eqref{hgh2}, we get 
\be\label{sol3} 
Z_3(n) = { 1 \over (n!)^2 } \sum_{ p \,\vdash n }  \Big( { n! \over \Sym ( p  ) }  \Big)^2  ( \Sym ( p  ))^3    
= \sum_{ p\, \vdash n } \Sym ( p ) \,, 
\qquad 
\Sym(p):= \prod_{i=1}^n (i^{p_i})(p_i!)
\ee
where the sum over $p= \{p_1 , p_2 , \cdots , p_n  \} $ is performed over all partitions of 
$n=\sum_i i p_i$.  The cardinality of a conjugacy class $T_p$ of $S_n$ with 
cycle structure determined by a partition $p$ is given 
by $|T_p|= n!/\Sym(p)$ (see Proposition \ref{prop:ncc} in the same appendix).

We can generate this sequence (using a GAP or Mathematica program, 
see GAP code 1 and Mathematica code 1, in Appendix \ref{app:gap}) and get 
\be
\label{SumSymMu} 
1, 4, 11, 43, 161, 901, 5579, 43206, 378360, 3742738, \dots 
\ee
This series is recognized in the OEIS website as A110143.  The same sequence also matches
with the counting of $n$-fold coverings of a graph  \cite{kwak}. This link will be clarified 
through the discussion vis permutation-TFT in Section \ref{sect:connect}. 

The number $Z_3(n)$ actually includes disconnected invariants. 
One can easily give a graphical representation to the
 first order terms:

-  $Z_3(1)=1$ consists in a single connected mass term (see Figure \ref{fig:tensinv} A) of the form 
\be
\sum_{i,j,k}\bar\Phi_{ijk}\Phi_{ijk}
\ee

- $Z_3(2)=4$ consists in 3 connected invariants (see Figure \ref{fig:tensinv} B), one of these is given by 
\bea
\sum_{i,i'}\bar\Phi_{i_1i_2i_3}\,\Phi_{i'_1i_2i_3}\,\bar\Phi_{i'_1i'_2i'_3}\, \Phi_{i_1i'_2i'_3}
\eea
and the 2 others are obtained by simple color permutation $1\to 2\to 3$,
plus one disconnected invariant  of the form 
\be
\Big(\sum_{i,j,k}\bar\Phi_{ijk}\Phi_{ijk}\Big)^2
\ee
This term is nothing but  twice a mass term (same as in Figure \ref{fig:tensinv} A above). 
Such disconnected invariant terms in higher rank Tensorial Group Field Theory framework should 
 be interesting since they appear as ``anomalous'' terms generated by the Renormalization Group flow \cite{BenGeloun:2011rc}.

\subsection{Connected invariants} 

To get the connected invariants, we can use the so-called plethystic logarithm (Plog) function
(for recent applications of this function in supersymmetric gauge theory and further references,   see \cite{PlethPro}).   This can be achieved in the following manner. 
 Define the generating function of the disconnected invariants as 
\be
Z_3 ( x ) = \sum_{ n =0}^{ \infty } Z_3 ( n ) x^n 
\label{z3x}
\ee
The Plog of $Z_3(x)$ is the function
\bea \label{plogz3}
\plog [ Z_3  ( x ) ] &=& \sum_{ k = 1 }^{ \infty }  { \mu ( k ) \over k } \log [ Z_3 [ x^k ] ] \label{plog3}\\\cr\cr
\mu(k) &=&
\left\{\begin{array}{cc}
0 & \text{if $k$ has repeated prime factors} \\
1 & k=1 \\
(-1)^{n} & \text{if $k$ is a product of $n$ distinct primes} \\
 \end{array} \right. 
\eea
where $\mu(k)$ is the so-called M\"obius function. 
This series can be expanded at finite order by  Mathematica 
(see Mathematica code 1 in Appendix \ref{app:gap}). We get the
following expansion at the lowest order as 
\be
x + 3x^2 + 7 x^3 + 26 x^4 +  97 x^5 + 624 x^6 + 4163 x^7 + \cdots 
\label{r3conX}
\ee
where $a_n$ the coefficient of $x^n$ gives now the number of
connected diagrams with $n$ black vertices (corresponding to $\bar \Phi$ ) 
and $n$ white vertices (for $\Phi$).    
This is again recognized in the OEIS as the series A057005 giving the number of conjugacy classes of subgroups of index $n$ 
in the free group of rank $2$. The first orders $Z_{3;\conex}(1)=1$,  $Z_{3;\conex}(2)=3$
and $Z_{3;\conex}(3)=7$ are represented graphically in  Figure \ref{fig:sss0}.  

\begin{figure}[h]\begin{center}
     \begin{minipage}[t]{.8\textwidth}\centering
\includegraphics[angle=0, width=9.5cm, height=6cm]{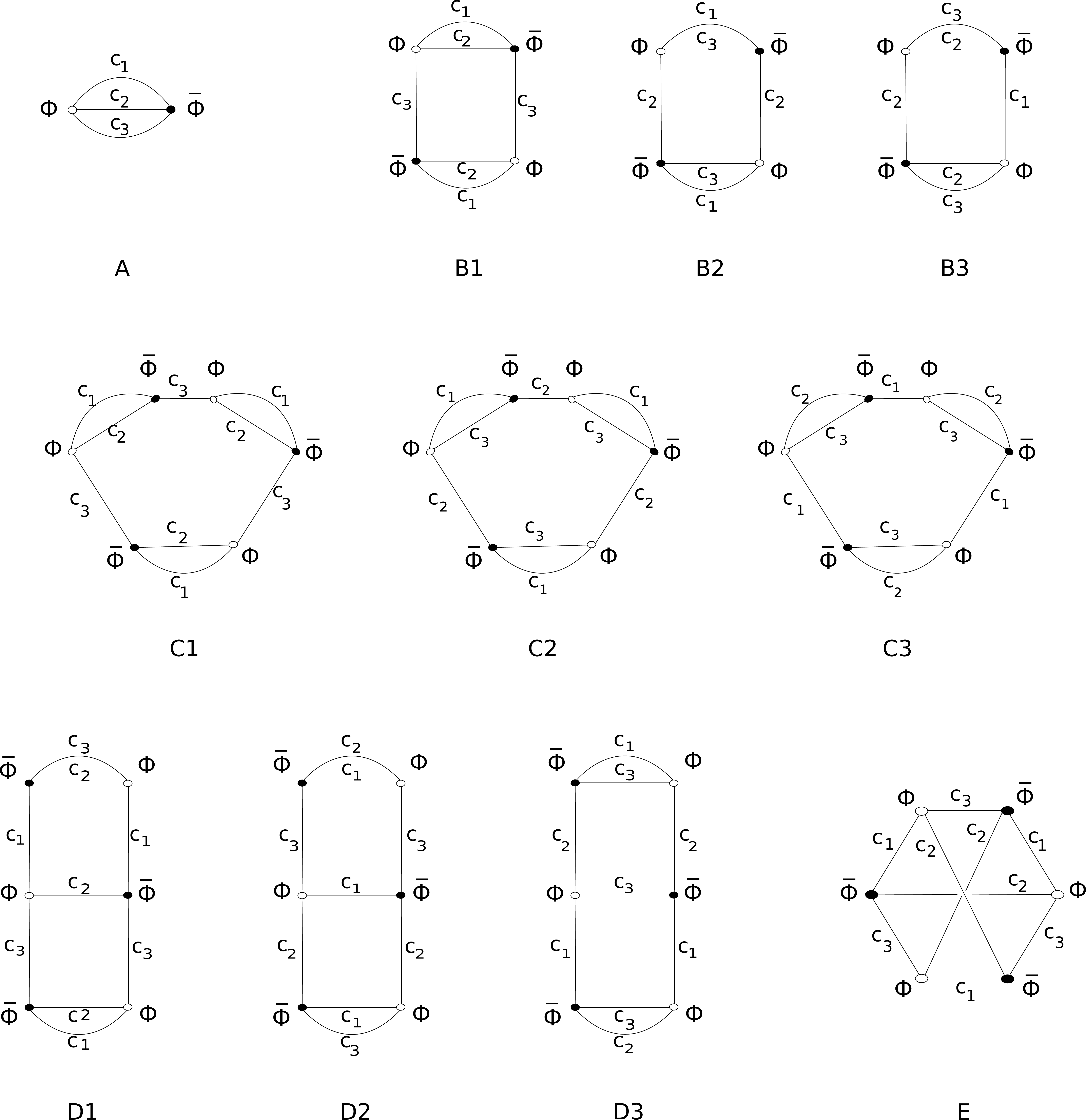}
\vspace{0.3cm}
\caption{ {\small The colored graphs associated with $Z_{3;\conex}(n)$: 
$n=1$, A; $n=2$, B1,B2 and B3; $n=4$, C1, C2, C3, D1, D2, D3 
and E}} \label{fig:sss0}
\end{minipage}
\end{center}
\end{figure}

Returning to a previous sequence, $Z_{3}(n)$ \eqref{SumSymMu} includes disconnected invariants. For the  first orders, $Z_3(2)=4$ includes B1, B2 and B3 as connected
objects plus a disconnected graph given by twice A; $Z_3(3)=11$,
contains all 7 connected graphs with 6 external legs which are C1, C2,\dots, E, drawn in Figure \ref{fig:sss0},
plus 4 other disconnected graphs given by the following combinations
of connected pieces: (A,A,A), (A,B1), (A,B2) and (A,B3) (we will keep 
that notation for disconnected components graphs).

\medskip 
\subsection{  Generalized rank $d$ case } 
 
For rank $d$ tensors, using $d-$tuples of permutations $(\sigma_1, \dots, \sigma_d)$ $\in (S_n)^{\times d}$ equivalent under the diagonal 
action  $\diag (S_n)$ such that 
\be
(\sigma_1, \dots, \sigma_d) \sim (\gamma_1\sigma_1\gamma_2, \dots, \gamma_1\sigma_d\gamma_2)
\ee
 following the same procedure and in adapted notations,  it is direct to obtain the number of tensor invariants made with $2n$  fields as
\bea\label{ncopdtens} 
Z_d ( n ) = \sum_{ p \,\vdash n }  ( \Sym ( p ) )^{ d-2} 
\eea
Given $d$ and $n$, this number can be evaluated by a GAP or Mathematica program (see GAP and Mathematica code 1, in Appendix \ref{app:gap}).  
This counting function $Z_d ( n ) $ has also been studied 
in connection with counting of $n$-fold coverings of the one-vertex graph with $(d-1)$ edges (which we denote by $F_{d-1}$, the flower graph with $d-1$ legs) which is equivalent to counting $n$-fold branched covers of the sphere with 
$d$ branch points \cite{kwak}.  The link between the tensor-invariant counting, which we related to  
the double coset $  \diag ( S_n ) \setminus S_n^{ \times d } / \diag ( S_n ) $, and the counting of 
covers will become clearer when we develop the permutation-TFT description in the next section.

For the $d=4$ case, the counting of invariants yields the sequence
\be
1,\, 8 ,\, 49,\,  681,\, 14721,\,  524137,\,  25471105, \,\dots 
\label{ran4}
\ee
for $n=1,2,3,4,5,6,7,\dots,$ respectively. 
For the case of  connected invariants
we use  the Plog function to  get, for $n=1,2,3,4,5,\dots,$ 
\be\label{r4conX}
1,\,7,\,41,\,604,\, 13753, \,\dots 
\ee
This is recognized as the A057006 sequence by OEIS or as the number of conjugacy  class of subgroups of index $n$ in the free group of rank 3. 
This sequence is also discussed the context of connected covers of $F_{d-1}$ 
 in \cite{kwak}.

\section{Tensor model invariants and permutation-TFTs }
\label{sect:connect}

In Section \ref{sect:Counting}, the  counting of 
tensor invariants was related to  the number of points in a  double coset. 
To calculate this we used a sum over group elements weighted by a delta function 
over the group (\ref{doubcosdelta}) to arrive at the formula (\ref{hgh2}). 
Such delta functions arise in a very simple physical construction, namely 
topological lattice gauge theory, where permutation groups play the role of 
gauge groups.  We give a brief review of this construction, and refer the reader to a more detailed 
review in  Section 5.1 of \cite{quivcal}  and the original literature \cite{DW,FHK}.
 Then we will show that 
the topological invariance of this lattice construction  illuminates the link between 
the counting of tensor invariants and the counting of branched covers of the 2-dimensional 
sphere.

\subsection{ Permutation TFTs - lightning review }\label{permTFTrev} 
 
On any cellular complex $ X $, one can define a partition function 
for a finite group $G$ by assigning a group element $g_{e}$
to each edge $e$ and to each plaquette $P$ a weight $w(g_{P})$,
where $g_P=\prod_{e \in P} g_e$. A most simple and natural 
choice independent of the plaquette size is given by 
\be
w(g_P) = \delta(g_P) = \left\{\begin{array}{c} = 1 \quad \hbox{ if} \quad  g_P={\rm id}\\
= 0 \quad \hbox{ otherwise }\end{array}\right.
\ee
The partition function in the model is given by 
 \be\label{defparTFT}
Z[ X ; \, G] = \frac{1}{|G|^V } \sum_{g_e} \prod_{P} w_P(\prod_{e\in P}g_e)
\ee 
where $V$ is the number of vertices in the cell deconposition. 
This theory is topological in the sense that it is invariant under
refinement of the cellular decomposition. We will be interested in cases where $G$ is taken to 
be the symmetric group $S_n$, of all permutations of $n$ objects. 
This simple topological field theory construction, with $n$ arbitrary,  
has a variety of applications in QFT combinatorics \cite{FeynCount,refcount,quivcal}. 

\begin{figure}[h]\begin{center}
     \begin{minipage}{.8\textwidth}\centering
\includegraphics[angle=0, width=3.5cm, height=2cm]{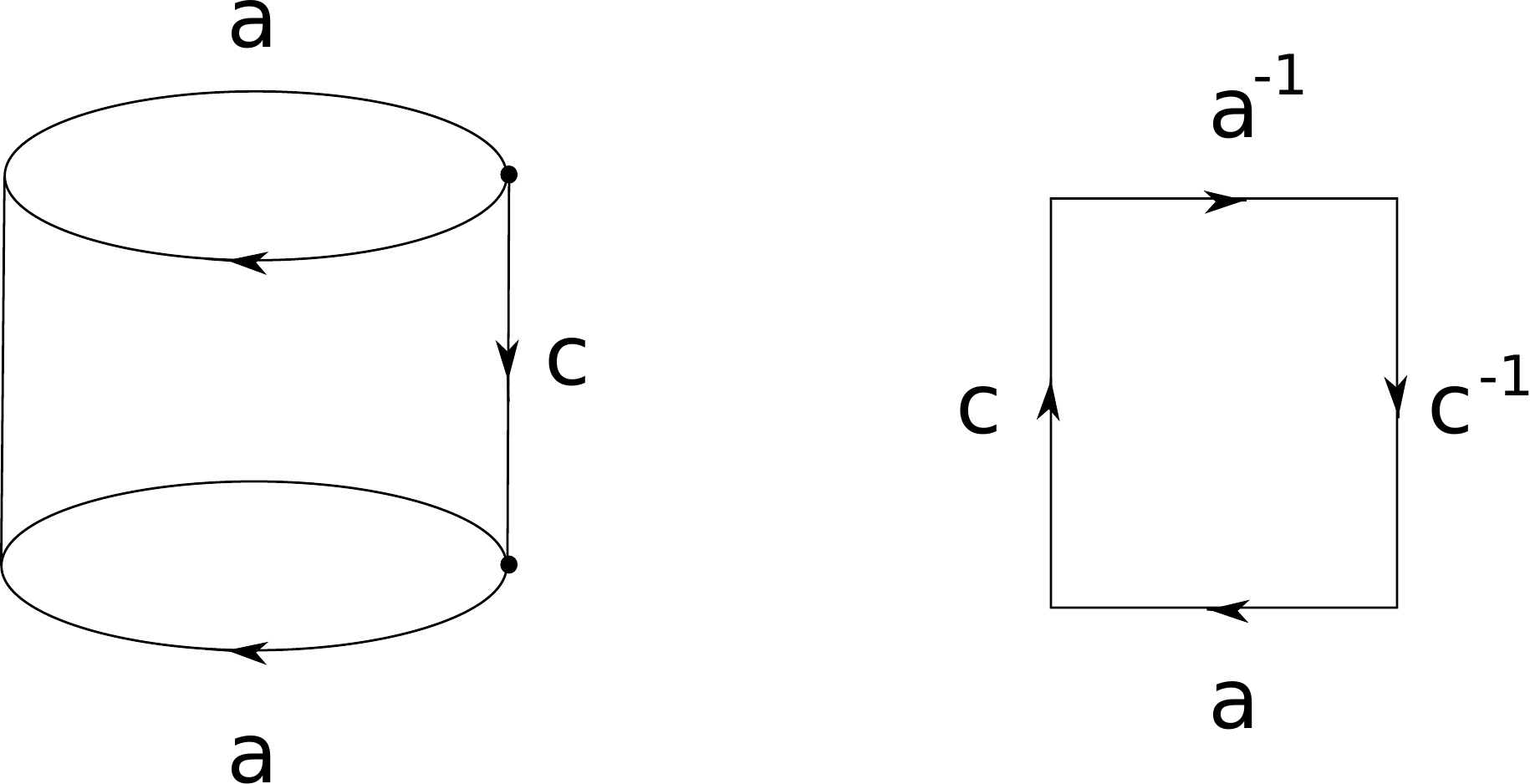}
\vspace{0.5cm}
\caption{ {\small Periodic torus and the plaquette action.}} \label{fig:pertor}
\end{minipage}
\put(-223,-15){A}
\put(-160,-15){B}
\end{center}
\end{figure}

Take the torus realized as a rectangle, with opposite sides identified (Figure \ref{fig:pertor}). 
This is a cell decomposition with a   single 0-cell, two 1-cells $a , c$ and a single 2-cell. 
Assign to each 1-cell 
a group element in $G$:  
\be\label{ac}
a \longrightarrow \s \,, \qquad 
c \longrightarrow \gamma
\ee
Thus the plaquette weight for the single 2-cell (plaquette) is
\be\label{plator}
w(g_P)=\delta(\gamma \s \gamma^{-1}\s^{-1})
\ee 
and the partition function is 
\bea\label{partTor} 
Z ( T^2  ;  S_n ) = {1 \over n! }  \sum_{ \sigma , \gamma \in S_n } \delta ( \gamma \sigma \gamma^{-1} \sigma^{-1} ) 
\eea
This partition function, for a topological space $X$, 
 counts equivalence classes of homomorphisms from $\pi_1 ( X) $ to $S_n$
 (weighted by the number of elements of $S_n$  which fix the homomorphism 
  under conjugation).   By a standard theorem of algebraic topology, this is equivalently counting 
  equivalence classes of   covering spaces of $X$ of degree $n$ (see e.g.  \cite{Hatcher}), counted with weight 
  equal to inverse of the order of the automorphism group of the cover).    
  The partition function (\ref{partTor}) thus counts $n$-fold covers of the torus and plays a role 
   in the string theory interpretation of two-dimensional YM theory \cite{GRT1,GRT2,CMR2}. 
   Given a cover, we can pick a generic point on the target space, label the inverse images 
    $\{ 1, \cdots , n \}$ and obtain permutations $\sigma , \gamma \in S_n $ as  we follow 
     the inverse images  of the 1-cells $a,c$  on the torus. The combination $ aca^{-1} c^{-1} $ 
     is a contractible path (shrinkable on the rectangle of Fig. \ref{fig:pertor}), so must give a trivial permutation of the sheets, which is enforced by  the delta function. 

\subsection{Toplogical invariance of  permutation-TFT : Double coset to conjugation equivalence }

We first start with  the rank 3 case and then generalize the ideas
to any rank $d$. The partition function $Z_3(n)$ can be written by applying 
Burnside's lemma (see Appendix \ref{app-group-basics}, Proposition \ref{prop:BL}) as 
\bea
Z_3 ( n ) &=&  { 1 \over n!^2  } \sum_{ \gamma_1 , \gamma_2 \in S_n } \sum_{ \sigma_1 , \sigma_2 , \sigma_3 \in S_n } 
\delta ( \gamma_1 \sigma_1 \gamma_2  \sigma_1^{-1} ) 
 \delta ( \gamma_1 \sigma_2 \gamma_2 \sigma_2^{-1} ) \delta ( \gamma_1 \sigma_3 \gamma_2 \sigma_3^{-1} ) 
 \label{gam1}
\eea  
Having seen the connection between sums over group delta functions and 
lattice TFTs, the natural question is : What topological space has a permutation-TFT 
partition function given by $Z_3(n)$ ? This allows us to see an emergence of geometry 
(more precisely topology at this stage, but see comment on holomorphic maps later)
directly from the structure of the counting problem.

\begin{figure}[h]\begin{center}
     \begin{minipage}[t]{.8\textwidth}\centering
\includegraphics[angle=0, width=12cm, height=3.2cm]{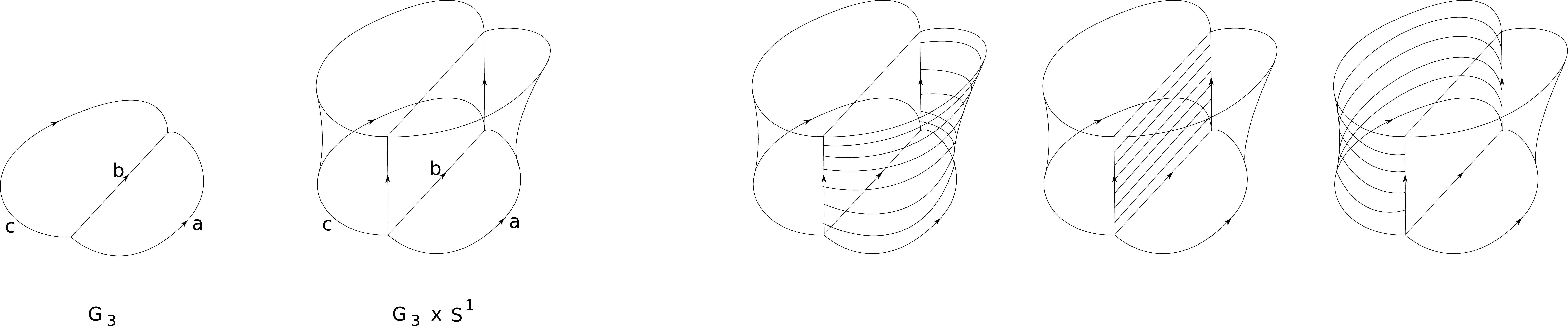}
\vspace{0.1cm}
\caption{ {\small $G_3$, $G_3 \times S^1$ and its different plaquettes (shaded)}} \label{fig:sep}
\end{minipage}
\put(-279,37){\scriptsize{$d$}}
\put(-250,65){\scriptsize{$d'$}}
\end{center}
\end{figure}

Consider the graph $G_3$ in  Figure \ref{fig:sep}, which has two vertices and 
three edges.  Next consider $G_3 \times S^1$, which  can be visualized as being 
obtained by evolving $G_3$ along a vertical time direction and then compactifying the time, 
which amounts to identifying the graph at the base of the Figure \ref{fig:sep} 
with the one at the top. The three 2-cells of this cell-complex are shaded. 
To do $S_n$ permutation-TFT on this complex, we assign 
\be
a \longrightarrow \s_1 \,, \qquad 
b \longrightarrow \s_2\,, \qquad
c \longrightarrow \s_3 
\ee
where the $\s_i \in S_n $. 
and we have two extra edges $d$ and $d'$ to which  we assign 
\be
d \longrightarrow \gamma_1 \,, 
\qquad 
d' \longrightarrow \gamma_2
\ee
with $ \gamma_i \in S_n $ . 
The partition function of this complex computed according to  (\ref{defparTFT}) as 
\bea 
Z ( G_3 \times S^1  ; S_n ) = { 1 \over n!^2 } 
\sum_{ \gamma_1 , \gamma_2 \in S_n } \sum_{ \sigma_1 , \sigma_2 , \sigma_3 \in S_n } 
\delta ( \gamma_1 \sigma_1 \gamma_2  \sigma_1^{-1} ) 
 \delta ( \gamma_1 \sigma_2 \gamma_2 \sigma_2^{-1} ) \delta ( \gamma_1 \sigma_3 \gamma_2 \sigma_3^{-1} ) 
\eea
We thus recognize that {\it the counting function for 3-index colored tensor invariants is 
the permutation-TFT partition function on  $G_3 \times S^1$ } : 
\be
\boxed{~~~~~~~  Z ( G_3  \times S^1  ; S_n )  = Z_3 ( n ) ~~~~~~~ } 
\ee
As observed in the lightning review, we can interpret this as counting covering spaces 
of $G_3 \times S^1$ - and  this is counting the covering spaces, with weight 
equal to inverse symmetry factor (see for example \cite{GRT2,CMR2} for explanation of this fact).  
 
The power of the permutation-TFT approach is that, not only, it exposes the 
geometry behind counting problems, but it also allows easy manipulations
of the delta functions, which often reveal  connections to other geometrical interpretations of 
the same counting problem.  In this case, we can use one the 
delta functions to solve for 
$\gamma_1 $ 
\bea 
Z_3(n) & = &   { 1 \over n!^2  } \sum_{ \gamma_2 } \sum_{ \sigma_1 , \sigma_2 , \sigma_3 \in S_n } 
 \delta (  \sigma_1 \gamma_2^{-1}  \sigma_1^{-1} \sigma_2 \gamma_2 \sigma_2^{-1} ) 
\delta ( \sigma_1 \gamma_2^{-1}   \sigma_1^{-1} \sigma_3 \gamma_2 \sigma_3^{-1} ) \cr 
&= & { 1 \over n! } \sum_{ \gamma \in S_n } \sum_{ \tau_1 , \tau_2  \in S_n } 
\delta ( \gamma \tau_1 \gamma^{-1} \tau_1^{-1} ) \delta ( \gamma \tau_2 \gamma^{-1} \tau_2^{-1} ) 
\label{gam2}
\eea
In the last line, we have defined $ \tau_1 = \sigma_1^{-1} \sigma_2 , \tau_2 = \sigma_1^{-1} \sigma_3 $, 
used the invariance of the $\sigma_2, \sigma_3 $ sums under this redefinition.  We also renamed $\gamma_2 \rightarrow \gamma$. 
 Now recalling Burnside's lemma again, we see that this is counting 
pairs $ ( \tau_1 , \tau_2 ) $ subject to the equivalence 
\bea\label{gammatau}  
 ( \tau_1 , \tau_2 ) \sim ( \gamma \tau_1 \gamma^{-1} , \gamma \tau_2 \gamma^{-1} ) 
\eea
Physically, these manipulations amount to starting from the equivalences 
\bea 
( \sigma_1 , \sigma_2 , \sigma_3 ) \sim ( \gamma_1 \s_1 \gamma_2 , \gamma_1 \s_2 \gamma_2 , \gamma_1 \s_3 \gamma_2)  , 
\eea
using the $\gamma_1 $  gauge symmetry : 
\bea 
( \s_1 , \s_2 , \s_3 ) \rightarrow ( 1 , \tau_1 \equiv  \s_1^{-1} \s_2 , \tau_2 \equiv \s_1^{-1} \s_3 ) 
\eea
After which the gauge equivalence $ \gamma_2 \rightarrow \gamma $ gauge symmetry 
becomes (\ref{gammatau}).  

\begin{figure}[h]\begin{center}
     \begin{minipage}[h]{.8\textwidth}\centering
\includegraphics[angle=0, width=5cm, height=2cm]{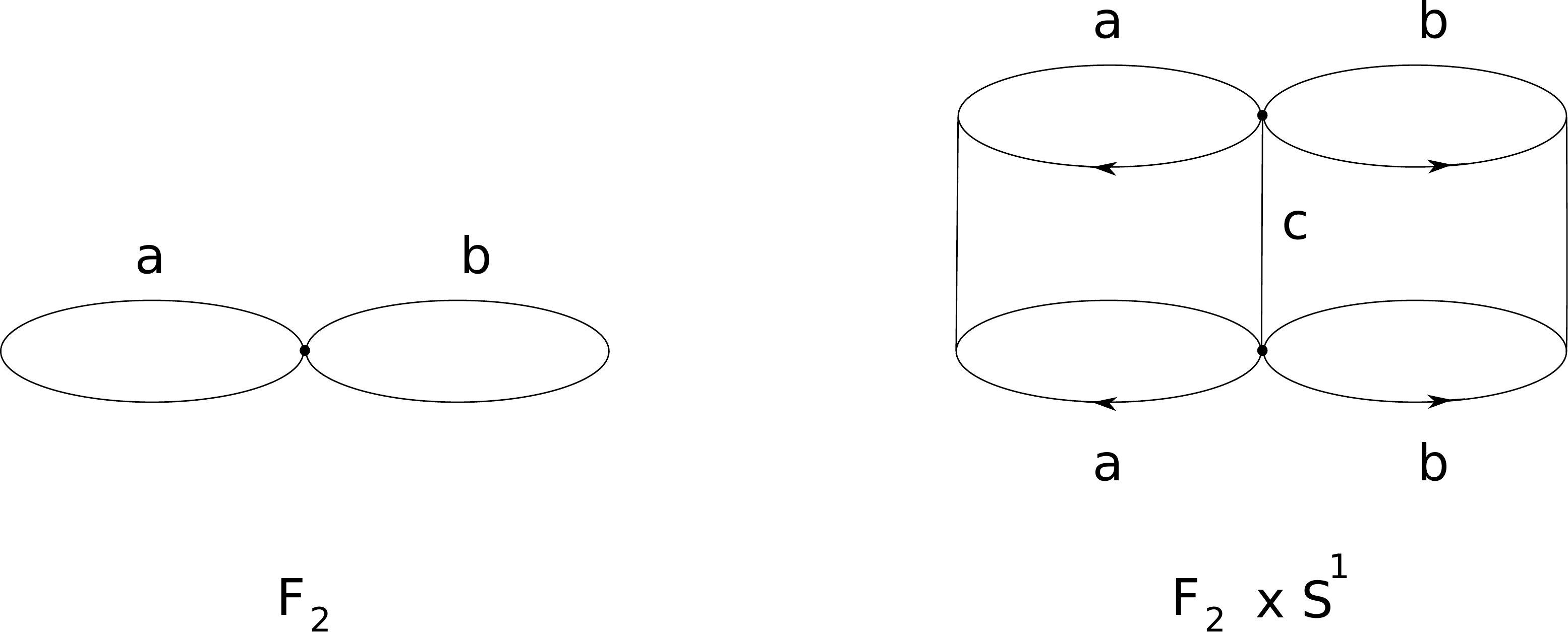}
\caption{ {\small The flower $F_2$ and its periodic lattice $F_2 \times S^1$}} \label{fig:flow}
\end{minipage}
\end{center}
\end{figure}

Now let us interpret the outcome geometrically. 
We observe that the expression (\ref{gam2})  coincides with 
 a permutation-TFT partition function for a simpler 
 cell-complex. This is $F_2 \times S^1$, where $F_2 $ is the Flower 
 graph, with a single vertex and two edges illustrated in Figure \ref{fig:flow}.
The flower  $F_2$  has a fundamental group made of two generators without any relations. 
 Consider the periodic flower $F_2 \times S^1$ as given in Figure \ref{fig:flow}.
  Opening $F_2 \times S^1$, we get, in the similar
way as \eqref{ac}, the following assignments
\be
a \longrightarrow \s_1 \,, \qquad 
c \longrightarrow \gamma\,, \qquad
b \longrightarrow \s_2 
\ee
and to the two different plaquettes present in the theory 
we assign a weight analogous to \eqref{plator} as
\be
w(g_{P_a}) =\delta(\gamma \s_1 \gamma^{-1}\s_1^{-1})\,,
\qquad 
w(g_{P_b}) =\delta(\gamma \s_2 \gamma^{-1}\s_2^{-1})
\ee
Thus, we identify the partition function of this $S_n$-TFT 
over the periodic cellular complex $F_2 \times S^1$ with our
previous counting: 
\be
Z ( F_2\times S^1;\, S_n ) = Z_3(n)
\ee
Now we have 
\bea\label{TwoCoveringInterps}  
 Z_3(n) = Z ( F_2\times S^1;\, S_n )  = Z ( G_3 \times S^1;\, S_n )
\eea
Since the  $S_n$-TFT  $ Z ( X  ; S_n ) $ simply counts homomorphisms 
$ \pi_1 ( X ) \rightarrow S_n $, the last equality is just the topological fact 
that 
\be
 \pi_1 ( F_2\times S^1 ) = \pi_1 (  G_3\times S^1  ) \,, 
\qquad  \pi_1 ( F_2 )  = \pi_1 (  G_3 )
\ee
In more physical terms, these relations give an example of the 
statement that the $S_n$-TFT is a topological field theory, with 
partition function invariant under a coarsening of the lattice which leaves the 
fundamental group invariant. The transformation leading from $G_3  \times S^1$
to $F_2 \times S^1$  shrinks the middle 2-cell in Figure \ref{fig:sep} thus identifying 
the two edges $d$ and $d'$. 

\subsection{ Conjugation equivalence, embedded bi-partite graphs, matrix models, branched covers } 

Let us return to the formulation of  the counting 
in terms of conjugation equivalence of the pair $ ( \tau_1 , \tau_2 ) $ which is 
expressed, via the Burnside lemma in (\ref{gam2}). We can manipulate this expression 
by introducing another permutation $\tau_0$  constrained by $ \tau_0 = ( \tau_1 \tau_2 )^{-1} $ 
\bea\label{triplesequivs}  
 Z_3 ( n ) & =&  { 1 \over n! } \sum_{ \gamma \in S_n } \sum_{ \tau_1 , \tau_2  \in S_n } 
\delta ( \gamma \tau_1 \gamma^{-1} \tau_1^{-1} ) \delta ( \gamma \tau_2 \gamma^{-1} \tau_2^{-1} ) \cr 
& = &   { 1 \over n! } \sum_{ \gamma \in S_n } \sum_{ \tau_0 , \tau_1 , \tau_2  \in S_n } 
\delta ( \gamma \tau_1 \gamma^{-1} \tau_1^{-1} ) \delta ( \gamma \tau_2 \gamma^{-1} \tau_2^{-1} ) 
\delta ( \tau_0 \tau_1 \tau_2 ) \cr 
& = &  { 1 \over n! } \sum_{ \gamma \in S_n } \sum_{ \tau_0 , \tau_1 , \tau_2  \in S_n } 
\delta ( \gamma \tau_0  \gamma^{-1} \tau_0^{-1} )
\delta ( \gamma \tau_1 \gamma^{-1} \tau_1^{-1} ) \delta ( \gamma \tau_2 \gamma^{-1} \tau_2^{-1} ) 
\delta ( \tau_0 \tau_1 \tau_2 ) 
\eea
In the last line  we introduced an extra delta function, implied by the 
ones already there, to make the formula more symmetric. 
We can recognize that this is counting, according to the Burnside lemma, 
triples of permutations $ \tau_0 , \tau_1 , \tau_2 $ obeying 
\bea\label{tripleONE}  
\tau_0 \tau_1 \tau_2 = 1 
\eea
More precisely, it is counting equivalence classes of these triples under the conjugation 
equivalence by $\gamma  \in S_n$  : $ \tau_i \sim \gamma \tau_i \gamma^{-1} $. 
We recognize in (\ref{tripleONE})   the group generated by three generators subject 
to one relation, which is the fundamental group of the two-sphere, with three 
punctures (equivalently 2-sphere with 3 discs removed).  Our counting function $Z_3 ( n )$  thus counts the number of equivalence classes 
of branched covers of the 2-sphere, with 3-branch points, each equivalence class being counted 
once\footnote{This is to be contrasted with the statement that $Z_3(n)$  counts equivalence classes 
of covers of $G_3 \times S^1$, not with weight one, but with weight equal to inverse automorphism 
group of these covers. As observed in \cite{refcount}, counting with  weight $1$  and with 
inverse automorphism are related via Burnside's lemma to introduction of an extra circle associated 
with $\gamma$.}. 
In two dimensions, branched covers are also holomorphic maps. 
These permutation triples thus  have a very rich mathematics : maps with three branch points  (which 
are often taken as $ 0,1,\infty $)  are called 
Belyi maps and are known to be definable over algebraic number fields \cite{Schneps}.  
Given such a
map, the inverse image of the interval $[0,1] $  gives an embedded bi-partite graph on the 
covering Riemann surface,
 where black vertices are inverse images of $1$ and white vertices are inverse images of 
$0$. These bi-partite graphs can be viewed as the large $N$ graphs of  matrix models
\cite{Gal,brown}.  Since branched covers in two dimensions are also holomorphic maps (defined by nice local 
equations which use the complex structure of the surfaces involved), this  has lead to 
investigations of links between these bi-partite graphs and  topological string theory \cite{Gal,Gop1,Gop2}. 
Our present observations relating the counting of 3-index tensor model invariants
to embedded bi-partite  graphs suggests that there may be surprising 
connections between these tensor models and  matrix models (and their associated 
gauge/string duals), with permutation-TFTs playing a key role. 
 We will  venture some more remarks in this direction in Section \ref{sec:discussions}. 

 The equation (\ref{triplesequivs}) was used as a starting point for  
refined counting of embedded bi-partite graphs in \cite{refcount}.  A very similar solving of delta functions, 
alongside Burnside's lemma, was used to uncover a surprising link between 
the counting of vacuum graphs in Quantum Electrodynamics and ribbon graphs \cite{FeynCount}.

\subsection{ General rank $d$   } 

Most  of the above discussion generalizes straightforwardly to higher rank. 
The counting of invariants built from $n$-copies of a rank $d$ colored tensor 
$ \Phi $ and $n$ copies of the conjugate $\bar \Phi$ is given by a function  $Z_d ( n )$ 
which coincides with the permutation-TFT partition function on $G_d \times S^1 $. 
$G_d$ is a graph with two vertices and $d$ edges. This partition function 
can be simplified to that of $F_{d-1} \times S^1 $, where $F_{d-1}$ is the flower 
graph with $d-1$ edges and a single vertex. 
\be
Z_d(n) = Z[F_{d-1} \times S^1;\, S_n] = Z[G_{d} \times S^1;\,S_n] \,,
\qquad 
\pi_1(F_{d-1}) =  \pi_1(S^2 \setminus d \;{\rm discs})
\ee
By introducing an extra permutation equal to the inverse of 
the $d-1$  permutations, we recognize the counting of
 equivalence classes of  branched covers of degree $n$ of
the sphere $S^2$ with $d$ branch points (each counted with weight $1$). 
The counting for the case of general $d$  is not known to us
 to have a simple matrix model realization, of the kind discussed above for $d=3$.

\section{Color-symmetrized counting of tensor invariants}
\label{sect:repack}

The simplest colored-tensor model e.g the Gaussian model, 
has a symmetry of permutations of the colors. It is natural to investigate the class
of interaction terms invariant under this symmetry. Here we will investigate 
the enumeration of these color symmetrized equivalence classes, express them 
in the language of permutations and obtain multi-variable generating functions 
for their counting.

\subsection{ Rank $d=3$ case} 

We start by the rank $d=3$ case
which will serve as a guiding non trivial situation. 
The color symmetrization can be achieved after imposing 
another type of equivalence now acting on the permutation triple as
\be
 ( \s_1  , \s_2 , \s_3 ) \sim ( \s_2  , \s_1 , \s_3 ) \sim ( \s_1 , \s_3 , \s_2 ) \sim \dots 
\label{sss123}
\ee 
As it stands, this problem turns out to nicely addressed using the group algebra $\IC(S_n)$ of $S_n$. 
Consider the element 
\be
[\s_{1}  \s_{2}  \s_{3} ]:=\sum_{\alpha \in S_3} \s_{\alpha(1)} \otimes \s_{\alpha(2)} \otimes
\s_{\alpha(3)} \; \in\;  \IC(S_n)^{\otimes 3} 
\label{symel}
\ee
Now we are investigating equivalence  classes given by  
\be
[\s_{1}  \s_{2}  \s_{3} ] \sim 
[\gamma_1^{\otimes 3} ][\s_{1}  \s_{2}  \s_{3} ]
[\gamma_2^{\otimes 3} ]
=\sum_{\alpha \in S_3} 
\gamma_1\s_{\alpha(1)}\gamma_2 \otimes \gamma_1\s_{\alpha(2)}\gamma_2 \otimes
\gamma_1\s_{\alpha(3)}\gamma_2
\ee
and we intend to find  $Z_{3;\,\col}(n)$ or the cardinal of  
\be
\diag(S_n) \ses \Sym(\IC(S_n)^{\otimes 3} )/\diag(S_n)
\ee 
with $\Sym(\IC(S_n)^{\otimes 3} )$ the group algebra generated by symmetric elements of the form \eqref{symel}. 

Using Burnside's lemma on $\IC(S_n)^{\otimes 3}$, we have
\bea
&&
Z_{3;\, \col}(n) = 
{\textstyle{\frac{1}{(3!)^2(n!)^2}}}\sum_{\gamma_1, \gamma_2 \in S_n}
\sum_{\s_i \in S_n}
\boldsymbol{\delta}\big([\gamma_{1}^{\otimes 3}]
[\s_{1} \s_{2}\s_{3}] [\gamma_{2}^{\otimes 3}][\s_{1} \s_{2}\s_{3}]^{-1}\big)
\crcr
&&:={\textstyle{\frac{1}{(3!)^2(n!)^2}}}\sum_{\gamma_1, \gamma_2 \in S_n}
\sum_{\s_i \in S_n}\sum_{\alpha, \beta\in S_3}
\delta(\gamma_{1}\,\s_{\alpha(1)}\gamma_2\,\s_{\beta(1)}^{-1})
\delta(\gamma_1\s_{\alpha(2)}\gamma_2\s_{\beta(2)}^{-1})
\delta(\gamma_1\s_{\alpha(3)}\gamma_2 \s_{\beta(3)}^{-1})\crcr
&& 
\label{inter}
\eea
We then use the same recipe  introduced before and integrate
one $\gamma$. Solving one delta function such that $\gamma_1 = \s_{\beta(1)}\gamma_2^{-1} \s_{\alpha(1)}^{-1}$, we rewrite \eqref{inter} as 
\bea
Z_{3;\, \col} (n)&=&{\textstyle{\frac{1}{(3!)^2(n!)^2}}}
\sum_{ \gamma_2 \in S_n}
\sum_{\s_i \in S_n}\sum_{\alpha, \beta\in S_3}
\delta(\s_{\beta(1)} \gamma_2^{-1} \s_{\alpha(1)}^{-1}\s_{\alpha(2)}\gamma_2\s_{\beta(2)}^{-1})\cr\cr
&&\qquad \qquad 
\delta(\s_{\beta(1)} \gamma_2^{-1} \s_{\alpha(1)}^{-1}\s_{\alpha(3)}\gamma_2 \s_{\beta(3)}^{-1})
\label{inter2}
\eea
We change dummy  variables $i \leftrightarrow \alpha^{-1}(i)$ so that 
\bea
Z_{3;\, \col} (n)&=&{\textstyle{\frac{1}{(3!)^2(n!)^2}}}
\sum_{ \gamma \in S_n}
\sum_{\s_i \in S_n}\sum_{\alpha, \beta\in S_3}
\delta(\s_{[\alpha^{-1}\beta](1)} \gamma^{-1} \s_{1}^{-1}\,\s_{2}\gamma\,\s_{[\alpha^{-1}\beta](2)}^{-1})\\ \cr
&&\qquad \qquad 
\delta(\s_{[\alpha^{-1}\beta](1)} \gamma^{-1} \s_{1}^{-1}\,\s_{3}\gamma\,\s_{[\alpha^{-1}\beta](3)}^{-1})
\label{inter3} 
\nn
\eea
Perform a last change in variable   $\alpha^{-1}\beta\to \beta$ and generate
\bea
Z_{3;\, \col} (n)&=&{\textstyle{\frac{1}{3!(n!)^2}}}
\sum_{ \gamma \in S_n}
\sum_{\s_i \in S_n}\sum_{\beta\in S_3}
\delta(\s_{\beta(1)} \gamma^{-1} \s_{1}^{-1}\,\s_{2}\gamma\,\s_{\beta(2)}^{-1})
\delta(\s_{\beta(1)} \gamma^{-1} \s_{1}^{-1}\,\s_{3}\gamma\,\s_{\beta(3)}^{-1}) \crcr
&=&
{\textstyle{\frac{1}{3!(n!)^2}}}
\sum_{ \gamma \in S_n}
\sum_{\s_i \in S_n}\Big\{ 
\delta(\s_{1} \gamma^{-1} \s_{1}^{-1}\,\s_{2}\gamma\,\s_{2}^{-1})
\delta(\s_{1} \gamma^{-1} \s_{1}^{-1}\,\s_{3}\gamma\,\s_{3}^{-1})  \crcr
 \qquad &+&
\delta(\s_{2} \gamma^{-1} \s_{1}^{-1}\,\s_{2}\gamma\,\s_{1}^{-1})
\delta(\s_{2} \gamma^{-1} \s_{1}^{-1}\,\s_{3}\gamma\,\s_{3}^{-1})  \crcr
 \qquad &+&
\delta(\s_{3} \gamma^{-1} \s_{1}^{-1}\,\s_{2}\gamma\,\s_{2}^{-1})
\delta(\s_{3} \gamma^{-1} \s_{1}^{-1}\,\s_{3}\gamma\,\s_{1}^{-1})  \crcr
 \qquad &+&
\delta(\s_{1} \gamma^{-1} \s_{1}^{-1}\,\s_{2}\gamma\,\s_{3}^{-1})
\delta(\s_{1} \gamma^{-1} \s_{1}^{-1}\,\s_{3}\gamma\,\s_{2}^{-1}) \crcr
 \qquad &+&
\delta(\s_{3} \gamma^{-1} \s_{1}^{-1}\,\s_{2}\gamma\,\s_{1}^{-1})
\delta(\s_{3} \gamma^{-1} \s_{1}^{-1}\,\s_{3}\gamma\,\s_{2}^{-1})  \crcr
\qquad &+&
\delta(\s_{2} \gamma^{-1} \s_{1}^{-1}\,\s_{2}\gamma\,\s_{3}^{-1})
\delta(\s_{2} \gamma^{-1} \s_{1}^{-1}\,\s_{3}\gamma\,\s_{1}^{-1}) \Big\}
\label{inter3}  \nn
\eea
These six terms come, respectively, from $ \beta = \{ { \rm id }  , (12) , (13) , (23), (132) , (123) \}  $.  
In each of the last three lines, $\s_3$ appears only once in at least one of the delta functions. 
So we can integrate these to be left  with a single delta function. For the last line, we also do
renaming of $\sigma_3 \rightarrow \sigma \gamma $ after the elimination of $\s_2$. The upshot is 
\bea 
Z_{3;\, \col}  ( n ) &=&  { 1 \over 6 n ! } 
\sum_{ \gamma \in S_n} \sum_{\s_2, \s_3 \in S_n}\delta(\gamma^{-1} \,\s_{2}\gamma\,\s_{2}^{-1})
\delta(\gamma^{-1} \,\s_{3}\gamma\,\s_{3}^{-1})  \crcr 
&+&   { 1 \over 2 n! }  \sum_{ \gamma \in S_n } \sum_{ \sigma \in S_n } \delta ( \gamma^2 \sigma \gamma^{-2}\sigma^{-1} ) \cr 
&+&   { 1 \over 3 n! } \sum_{ \gamma  , \sigma \in S_n  } \delta ( \gamma^3 \sigma^3 ) 
\eea
We know how to calculate the first sum in terms of a sum over partitions. 
We should be able to derive something similar for the last two terms. As a first step, 
we write
\be
Z_{3;\, \col } (n ) 
=  { 1 \over 6 n! } \sum_{ p  \vdash n }   \Sym ( p  ) 
   +  { 1 \over 2 n! }  \sum_{ \gamma \in S_n } \sum_{ \sigma \in S_n } 
\delta ( \gamma^2 \sigma \gamma^{-2}\sigma^{-1} ) + { 1 \over 3 n! } \sum_{ \gamma  , \sigma \in S_n  } \delta ( \gamma^3 \sigma^3 ) 
\label{s1s2s3}
\ee
Let us write this as 
\bea 
Z_{3; \col } ( n ) = { 1 \over 6 } S^{(3)}_{ [1^3]}  ( n ) + { 1 \over 2}  S^{(3)}_{[2,1]} ( n ) +
 { 1 \over 3 } S^{(3)}_{ [3]}  ( n ) 
\eea
where the superscript indicates that this is the $d=3$ case, while 
the subscript is a partition of $3$ corresponding to the conjugacy class of $\alpha$ 
which gives rise to the relevant term.  We record here our most effective formulae for each term 
\bea\label{Eff-formulae}  
 S^{(3)}_{ [1^3]}   ( n )  & = &  \sum_{ p \vdash n } \Sym ( p )  = 
\sum_{ p \vdash n } \prod_{i=1}^n (i^{\mu_i})(\mu_i!)   \cr 
 S^{(3)}_{[2,1]} ( n )  & = & \sum_{ p \vdash n }    \hbox{ Coefficient }  [  Z^{(2)} (  t , \vec x) , t^n x_1^{p_1} x_2^{p_2}\dots x_n^{p_n}  ] \times\big[  \prod_{i=1}^n i^{p_i} p_i!  \big]   \cr 
S^{(3)}_{ [3]}  ( n ) 
 &=&  \sum_{ p \vdash n }  (\hbox{Coefficient } [  Z^{(3) } ( t , \vec x )  , t^n \prod_i x_i^{p_i} ]   )^2 \times \big[\prod_i i^{p_i } p_i! \big]
\eea
The derivation of $ S^{(3)}_{ [1^3]}   ( n )  $  was explained earlier (\ref{sol3}).
  The formulae for $ S^{(3)}_{[2,1]} ( n ) $ 
and $S^{(3)}_{ [3]}  ( n ) $  in terms of multi-variable generating functions 
are explained and  derived as \eqref{S2nKeyFormula} and \eqref{S3nKeyFormula}  in Appendix \ref{app:s2n}. 
These formulae can be evaluated to high orders using  Mathematica
(see Mathematica code 2 in Appendix \ref{app:gap}). 
The result for $ S^{(3)}_{[2,1]} ( n )$  is\footnote{This is recognized as the sequence A082733
by OEIS, and described there  as the sum of all entries in the character table of $S_n$.}  
\bea 
1,2,5,13,31,89,259,842,2810, \cdots 
\eea
The sequence $ S^{(3)}_{ [3]}  ( n )$ evaluates in the same way as (see Appendix \ref{app:s2n} and Mathematica code 2 in Appendix \ref{app:gap})
\bea 
1,1,2,4,5,13,29,48,114,301 
\eea
Adding all these up with the right coefficients, we get 
\be\label{z3colconX}
1, 2, 5, 15, 44 , 199, 1069, \dots 
\ee
Note that the summands can be fractional, but the sum is integral. 
This is the disconnected case. Having a closer look at Figure \ref{fig:sss0},  we can associate the graphs to the first orders, $Z_{3; \col}(1)=1$ is simply the class given by A; for $Z_{3; \col}(2)=2$, there are two classes
of graphs: the first is given by a disconnected graph formed by twice (A,A), and the second class is formed by the three remaining B1, B2, B3
which are indeed form a closed set under the $S_3$ operations of permuting the three  colors.
 Now $Z_{3; \col}(3)=5$ is generated
by  $\{$(A,A,A)$\}$ (disconnected), {\bf AB}=$\{$(A, Bi), i=1,2,3$\}$ (disconnected),  {\bf C}=$\{$C1, C2, C3$\}$, {\bf D}=$\{$D1, D2, D3$\}$ and the last class given by {\bf E}=$\{$E$\}$.

\begin{figure}[h]\begin{center}
     \begin{minipage}[t]{.8\textwidth}\centering
\includegraphics[angle=0, width=6cm, height=2cm]{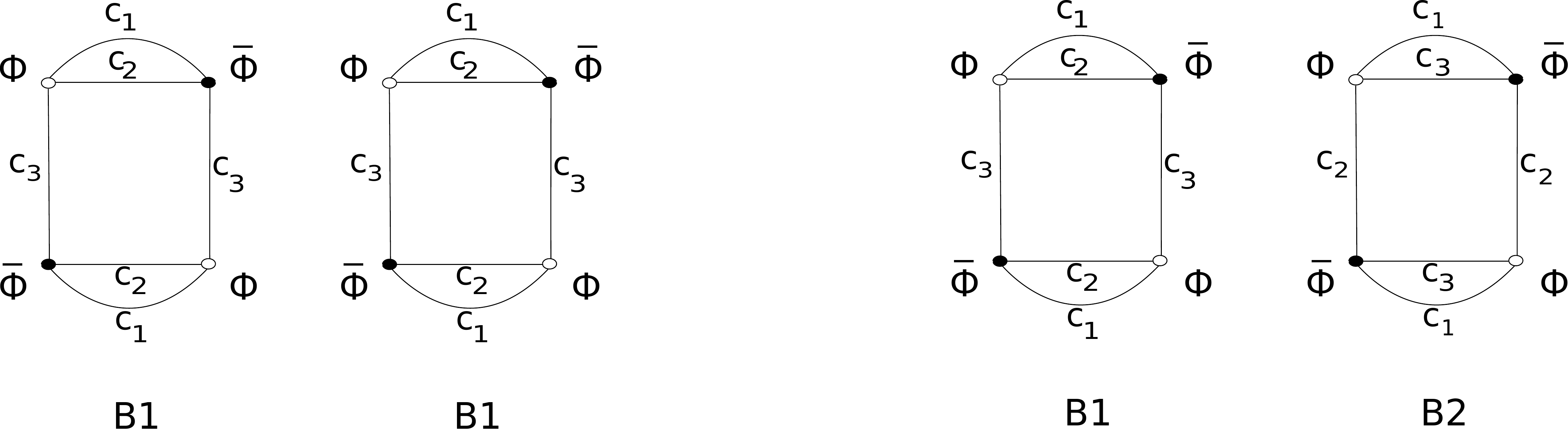}
\vspace{0.1cm}
\caption{ {\small Non equivalent disconnected graphs}} \label{fig:plog}
\end{minipage}
\end{center}
\end{figure}

\begin{figure}[h]\begin{center}
     \begin{minipage}[t]{.8\textwidth}\centering
\includegraphics[angle=0, width=10cm, height=6cm]{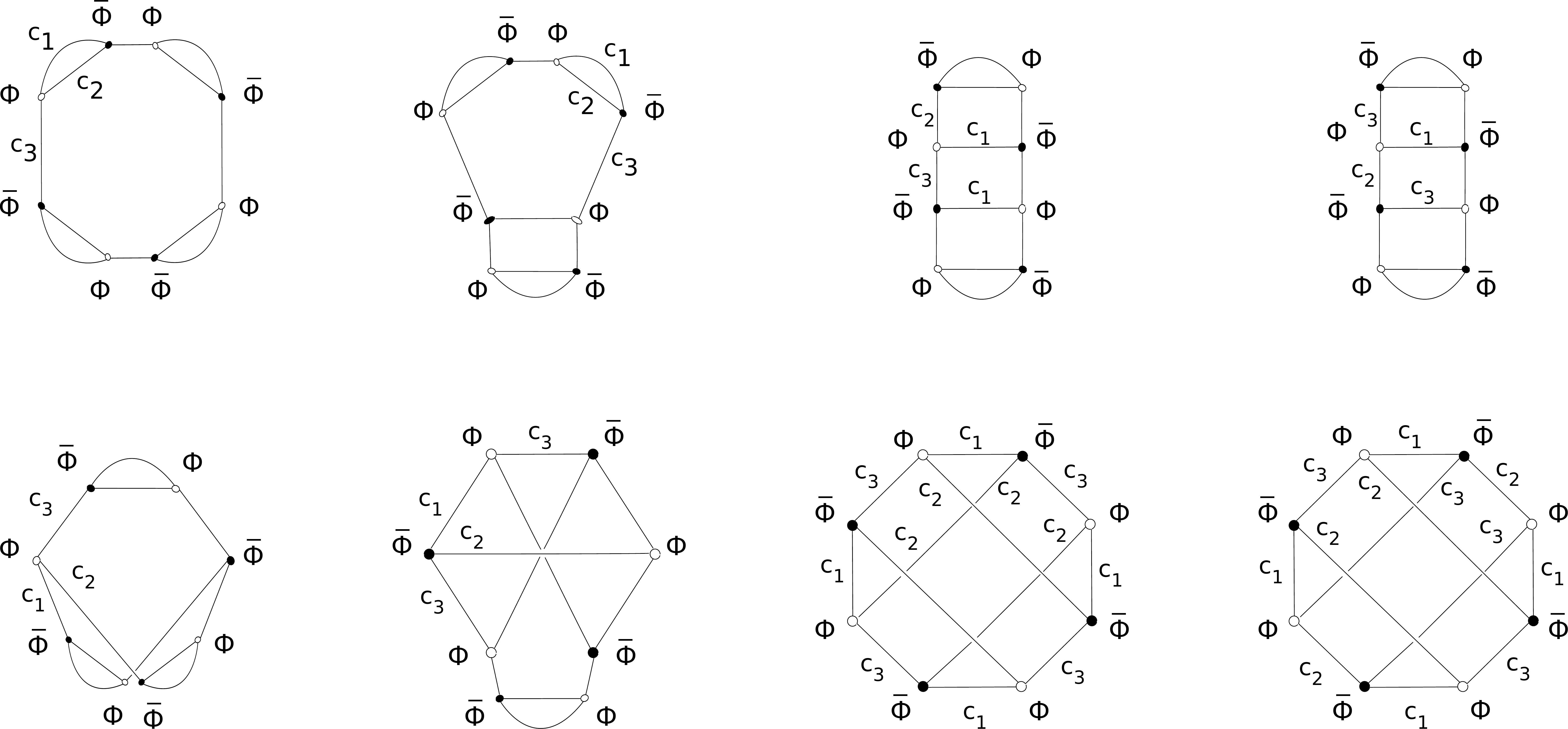}
\vspace{0.1cm}
\caption{ {\small Rank 3 colored symmetric connected invariants at order $n=4$}} \label{fig:n4}
\end{minipage}
\end{center}
\end{figure}

 It turns out that the Plog does not give the correct relation between 
 connected and disconnected for this color-symmetrized counting. 
For instance, at order $n=4$ (graph with 8 legs), the Plog gives 9. 
This means that it has subtracted 6 classes (from the initial 15 classes) regarded as disconnected.
 Now, from the case $n=3$, we can observe directly that these classes can be organized as follows: 
3 disconnected graphs are formed by  (A, {\bf C}), (A, {\bf D}) and (A, {\bf E}); another case is given by twice a copy
of A plus a connected piece with four legs, which gives (A, {\bf AB}); 
then we must also include the graph made with four copies of A, which is  
(A,A,A,A). That yields 5 cases already out of the 6. So the remaining
disconnected graph would be the one formed by twice a graph 
made with 4 legs (a double copy of Bi, i=1,2,3, see Figure \ref{fig:sss0}). 
However, in the latter category of disconnected objects, the class obtained by the disjoint union of graphs Bi denoted by $\{$(Bi, Bi)$\}$ and the one $\{$(Bi, Bj)$, i\neq j\}$ (see Figure \ref{fig:plog}) are not equivalent under \eqref{sss123}. Thus, the ordinary Plog of the disconnected series does not 
give the correct answer. It would be interesting to work out an analog of the Plog formula for this case
of color-symmetrized counting of invariants. 
A GAP program can however generate the sequence of connected graphs (see GAP code 1, in Appendix \ref{app:gap}). One finds
\be\label{r3colorconX}
1,1,3,8,24,72
\ee
The case $n=4$ giving $Z_{3;\,\col}^{\conex}(n=4)=8$ has been illustrated in Figure \ref{fig:n4}. 

\medskip 

\subsection{  Rank $d=4$ case.} 

The color symmetrization 
here can be implemented by the equivalence of the $d$--tuples
\be
 ( \s_1  ,\dots,   \s_d ) \sim ( \s_{\alpha(1)}  , \dots , \s_{\alpha(d)} )\,,
\qquad \forall \alpha \in S_d 
\ee 
Using now the group algebra $\IC(S_n)$ of $S_n$, we consider the element 
\be
[\s_{1}  \dots  \s_{d} ]:=\sum_{\alpha \in S_d} \s_{\alpha(1)} \otimes \s_{\alpha(2)} \otimes \dots \otimes \s_{\alpha(d)} \; \in\;  \IC(S_n)^{\otimes 3} 
\label{symeld}
\ee
which leads us to the search of equivalent classes such that 
\be
[\s_{1}  \dots  \s_{d} ] \sim 
[\gamma_1^{\otimes d} ][\s_{1}  \dots  \s_{d}  ]
[\gamma_2^{\otimes d} ]
=\sum_{\alpha \in S_3} 
\gamma_1\s_{\alpha(1)}\gamma_2 \otimes \dots  \otimes
\gamma_1\s_{\alpha(d)}\gamma_2
\ee
This is counting the points of  
\be
\diag(S_n) \ses \Sym(\IC(S_n)^{\otimes d} )/\diag(S_n)
\ee 
with $\Sym(\IC(S_n)^{\otimes d} )$ the group algebra
generated by symmetric elements of the form \eqref{symeld}. 

Burnside's lemma on $\IC(S_n)^{\otimes d}$ allows us to write
\be
Z_{d;\, \col}(n) ={\textstyle{\frac{1}{(d!)^2(n!)^2}}}\sum_{\gamma_1, \gamma_2 \in S_n}
\sum_{\s_i \in S_n}\sum_{\alpha, \beta\in S_d}
\delta(\gamma_{1}\,\s_{\alpha(1)}\gamma_2\,\s_{\beta(1)}^{-1})
\dots
\delta(\gamma_1\s_{\alpha(d)}\gamma_2 \s_{\beta(d)}^{-1})
\label{interd}
\ee
Integrating $\gamma_1$, $\gamma_1 = \s_{\beta(1)}\gamma_2^{-1} \s_{\alpha(1)}^{-1}$, \eqref{interd} re-expresses as  
\bea
Z_{d;\, \col} (n)&=&{\textstyle{\frac{1}{(s!)^2(n!)^2}}}
\sum_{ \gamma_2 \in S_n}
\sum_{\s_i \in S_n}\sum_{\alpha, \beta\in S_d}
\delta(\s_{\beta(1)} \gamma_2^{-1} \s_{\alpha(1)}^{-1}\s_{\alpha(2)}\gamma_2\s_{\beta(2)}^{-1}) \dots \cr\cr
&&\qquad \qquad \dots 
\delta(\s_{\beta(1)} \gamma_2^{-1} \s_{\alpha(1)}^{-1}\s_{\alpha(d)}\gamma_2 \s_{\beta(d)}^{-1}) 
\label{inter2d}
\eea
Changing variables as $i \leftrightarrow \alpha^{-1}(i)$ and  performing $\alpha^{-1}\beta\to \beta$  generate 
\bea
Z_{d;\, \col} (n)&=&{\textstyle{\frac{1}{(d!)^2(n!)^2}}}
\sum_{ \gamma \in S_n}
\sum_{\s_i \in S_n}\sum_{\alpha, \beta\in S_d}
\delta(\s_{[\alpha^{-1}\beta](1)} \gamma^{-1} \s_{1}^{-1}\,\s_{2}\gamma\,\s_{[\alpha^{-1}\beta](2)}^{-1}) \dots\cr \cr
&&\qquad \qquad \dots
\delta(\s_{[\alpha^{-1}\beta](1)} \gamma^{-1} \s_{1}^{-1}\,\s_{d}\gamma\,\s_{[\alpha^{-1}\beta](d)}^{-1})
\label{inter3d} \\ 
&=&{\textstyle{\frac{1}{d!(n!)^2}}}
\sum_{ \gamma \in S_n}
\sum_{\s_i \in S_n}\sum_{\beta\in S_d}
\delta(\s_{\beta(1)} \gamma^{-1} \s_{1}^{-1}\,\s_{2}\gamma\,\s_{\beta(2)}^{-1})\dots 
\delta(\s_{\beta(1)} \gamma^{-1} \s_{1}^{-1}\,\s_{d}\gamma\,\s_{\beta(d)}^{-1})\crcr
&& \nn
\eea
We now specialize to the case $d=4$. Expanding the last sum $\sum_{\beta\in S_d}$, one gets after some algebra:
\bea\label{z4scdeltas} 
Z_{4;\, \col}  ( n ) &=&  { 1 \over 24 n ! } 
\sum_{ \gamma \in S_n} \sum_{\s_i \in S_n}
\delta(\gamma^{-1} \,\s_{2}\gamma\,\s_{2}^{-1})
\delta(\gamma^{-1} \,\s_{3}\gamma\,\s_{3}^{-1})
\delta(\gamma^{-1} \,\s_{4}\gamma\,\s_{4}^{-1})  \crcr 
&& ~~~ + { 1 \over 4 n! }  \sum_{ \gamma \in S_n } \sum_{ \s_i  \in S_n } \delta( \gamma\s_{1}\gamma^{-1}\s_{1}^{-1} )\delta ( \gamma^2 \s_2 \gamma^{-2}\s_2^{-1} ) \cr 
&& ~~~  + { 1 \over 3 n! }\sum_{ \gamma \in S_n} \sum_{\s \in S_n}
 \delta(   \gamma^{3}\,\s \gamma^{-3} \s^{-1}  )  \cr 
&& ~~~  + { 1 \over 8 n! }\sum_{ \gamma \in S_n} \sum_{\s_i \in S_n}
\delta(\s_1^2\gamma^2\,)
\delta(\gamma^{2} \s_2 \gamma^{-2}\,\s_{2}^{-1} ) \crcr
&& ~~~ + { 1 \over 4 n! }  \sum_{ \gamma \in S_n } \sum_{ \s  \in S_n } \delta(\s^4 \gamma^{4}) 
\eea
These 5 terms come, respectively, from the conjugacy classes 
represented by 
$\{{ \rm id } ,(12) , (123)$ $(12)(34) , (1234)  \}$.  
As above, the first sum computes to a sum over partitions already known
from  \eqref{ran4}. Let us denote: 
\be
Z_{4; \col} (n) = 
 { 1 \over 24 } S^{(4)}_{[1^4]} ( n ) + { 1 \over 4}  S^{(4)}_{[2,1^2]} ( n ) 
+ { 1 \over 3 } S^{(4)}_{[3,1]} ( n ) 
+ {1 \over 8}  S^{(4)}_{[2^2]} (n) + { 1 \over 4}   S^{(4)}_{[4]} ( n )
\label{s1ps2p}
\ee
where, as in the rank 3 case, we can label each sum by a 
subscript giving by a particular partition of $d=4$. 
In Appendix \ref{app:s2n} we manipulate these delta functions to 
arrive at expressions as sums over symmetry factors of partitions 
or in terms of multi-variable generating functions. We summarize the key formulae  (see Appendix \ref{app:s2n}, \eqref{s431},
\eqref{s44}, \eqref{s4212} and \eqref{s422} for more details) : 
\bea\label{Eff-forms-d4}
S^{(4)}_{[1^4]} ( n )  & = &  \sum_{ p \vdash n } ( \Sym ~ p )^2 
 \cr
S^{(4)}_{[2,1^2]} ( n )  & = &  \sum_{ p \vdash n } \big[\prod_{ j =1 }^{ \lfloor { n \over 2 } \rfloor } (2j)^{2  p_{4j} } ( 2 p_{4j} )!\big]\big[ \prod_{ j=0}^{ \lfloor { n \over 2 } \rfloor } ( 2j+1)^{ p_{2j+1} + 2 p_{ 4j+2} }   ( p_{2j+1} + 2 p_{ 4j+2} ) !\big] 
\cr 
 S^{(4)}_{[3,1]} ( n )  & = & \sum_{ p \vdash n }    \hbox{ Coefficient }  [  Z^{(3)} (  t , \vec x ) , t^n x_1^{p_1} x_2^{p_2}\dots x_n^{p_n}  ] \times \big[  \prod_{i=1}^n i^{p_i} p_i!  \big]
   \cr 
 S^{(4)}_{[2^2]} (n) & = &  \sum_{ p \vdash n }     \Big( \hbox{Coefficient}[  Z^{(2)} (  t , \vec x ) , t^n x_1^{p_1} x_2^{p_2}\dots x_n^{p_n}  ]  \, \Sym(p)\Big)^2 
 \cr 
 S^{(4)}_{[4]} ( n ) & = &  S^{(4)}_{[4]} ( n ) =  \sum_{ p \vdash n }  \Big(\hbox{Coefficient} [Z^{(4)} ( t , \vec x )  , \prod_i x_i^{p_i} ]  \Big)^2 \times \big[\prod_i i^{p_i } p_i!  \big] 
\cr 
&& 
\eea
With these formulae in hand, we can generate the sequences to high order with Mathematica. 
Direct evaluation of the delta functions with GAP at low orders agrees with these generating functions 
but at high orders calculation with the help of  (\ref{Eff-forms-d4})  is the only practical option. 
The sequences $S^{(4)}_{[\cdot]} (n ) $ can be computed with Mathematica to give (see Appendix \ref{app:s2n} and Appendix \ref{app:gap}, 
Mathematica code 2, 3  and 4 for further details)
\bea 
S^{(4)}_{[2,1^2]} : \qquad &&
1, 4, 15, 83, 385, 2989, 20559, \dots  \cr \cr
S^{(4)}_{[3,1]}: \qquad &&
1, 2, 4, 12, 27, 103, 391, \dots\cr\cr
S^{(4)}_{[2^2]} : \qquad &&
1,4,17,105,685, 5825,54013, \dots
\cr \cr
 S^{(4)}_{[4]}: \qquad && 
1,2,3,11,27,93, 233,\dots
\eea
Combining these sums yields 
\be
1, 3, 10, 69, 811, 23372, 1073376, \dots  
\ee
This sequence corresponds to the disconnected case. The connected case sequence
can be obtained with a GAP program extending 
the rank $d=3$ case as given in Appendix \ref{app:gap}.

\section{Counting tensor invariants without color} 
\label{sect:other}

We address here countings of invariants for tensors without color, which are  the 
tensor models of more traditional interest.  In the first case, the tensor field $ \Phi_{ i_1 \cdots i_d } $ 
will have $d$ indices and we will allow contraction of any $i_a$  with any of the $d$ indices of 
$\bar \Phi_{ j_1 \cdots j_d } $. In the second case, there will again be no restriction on which $j$ 
a given $i$ can contract with, but the tensor field will be symmetric under $S_d$ 
permutations of its indices. We will have a family of counting problems 
for each integer $n$ corresponding to the number of $\Phi$ and $\bar \Phi$ fields.

\subsection{Invariants without color: general tensors} 

This is equivalent to count invariants in $\Sym_n  ( V^{\otimes d } )^{\otimes n }   \otimes  \Sym_n 
( \bar V^{\otimes d } )^{\otimes n }   $ under a diagonal $U(N)$ action. 
The $\Sym_n$  indicates the symmetrization of the $n$ copies, which arises 
from the fact that the $n$ copies of $\Phi$ and the  $n$ copies of $\bar \Phi$ 
can be permuted without changing the invariant. 
 The unitary group acts as 
\be
U^{\otimes n d  } \otimes \bar U^{ \otimes nd } 
\ee 
on  $  (V^{\otimes d } )^{\otimes n } \otimes ( \bar V^{\otimes d } )^{\otimes n } $ 
which descends to an action on the symmetrized subspaces. 
The contractions are given by permutations $ \sigma  \in S_{dn} $
(mixing all $dn$ indices)
and the equivalences that we seek   are encoded in  
\bea 
\sigma \sim \gamma_1 \,\sigma \,\gamma_2 
\eea
where  $ \gamma_1 , \gamma_2 \in S_n =  \diag ( S_n^{\times d } ) \subset S_{dn} $ (this is the embedding of $S_n$ in $S_{dn}$). 
Equivalently we are counting points in the double coset 
\bea 
 S_n \ses S_{dn } / S_n 
\eea
We can again use the formula 
\bea 
\cZ_{d;\, \ncol}(n) =  \frac{1}{(n!)^2}\sum_{ C } ( Z^{S_n \rightarrow S_{dn}}_C )^2\;  \Sym (C )
\eea
where the sum is over conjugacy classes of $S_{dn}$. 
For a given conjugacy class $C$, $Z^{S_n \rightarrow S_{dn}}_C$
counts the number of elements $(\s,\dots, \s)$ in $\diag(S_n^{\times d})$ that is in $C$. For $(\s,\dots, \s)$ to be in $C$, $C$ must have
a cycle structure of $d \,\times$ the cycle structure of $\s$. The latter
is entirely determined by partition of $n$ so that 
$ \Sym (C )=\prod_i i^{ d p_i } ( d p_i ) !$ (see Proposition \ref{prop:ncc},
in Appendix \ref{app-group-basics}). We finally get 
\be
\cZ_{d;\, \ncol}(n)  = \sum_{p \vdash n }  \frac{1}{(\Sym(p))^2}   \prod_i i^{ d p_i } ( d p_i ) !  
=  \sum_{ p \vdash n }    \prod_i{  i^{ (d-2)  p_i } ( d p_i ) ! \over  (p_i!)^2 } 
\ee
A Mathematica program allows to compute this (see Appendix \ref{app:gap}, Mathematica code 5). Doing this for $d=2$ (matrix models) we get the sequence  
\be 
2, 8, 26, 94, 326, 1196, \cdots 
\label{noncol2} 
\ee
This sequence is recognized by the OEIS as A067855 or 
the squared length of sum of $s_p^2$, where $s_p$ is a Schur function and $p$ ranges over all partitions of $n$. 

For $d=3$, we get 
\be 
6, 192, 10170, 834612, 90939630, 12360636540,  \cdots 
\label{noncol3}
\ee
This corresponds to the disconnected case. The Plog function can 
generate the connected situation along the lines 
\eqref{z3x}, \eqref{plogz3} and \eqref{r3conX} (Mathematica code 1 see Appendix \ref{app:gap}).

\subsection{ Invariants without color: symmetric tensors}

Consider a  complex  symmetric tensor $\Phi$ of rank $d$, such that  
\be 
\Phi_{ i_1 i_2 i_3 \dots i_d} = \Phi_{i_{\rho(1)}i_{\rho(2)} \dots i_{\rho(d)} }\,,\qquad \rho \in S_d
\ee
We want to know the number $\cZ_{d;\,\sym}(n)$ of bi-partite graphs that one can build by contracting $n$ copies of $\Phi$ (seen as vertices 
of valence $d$) with $n$ copies of $\bar\Phi$.  
In terms of a traditional invariant theory question, we are counting invariants of 
$U(N)$ acting on $\Sym_n ( ( \Sym_d ( V^{\otimes d } ) ) ^{\otimes n } \otimes  
\Sym_n ( ( \Sym_d ( \bar V^{\otimes d } )) ^{\otimes n } ) $. The $S_d$ symmetrization 
implicit in $\Sym_d $ comes from having symmetric tensors. The $S_n$ symmetrizations 
come from having $n$ copies of the same $\Phi$ and $n$ of the same $\bar \Phi$. 
 
\begin{figure}[h]\begin{center}
     \begin{minipage}[t]{.8\textwidth}\centering
\includegraphics[angle=0, width=5cm, height=2cm]{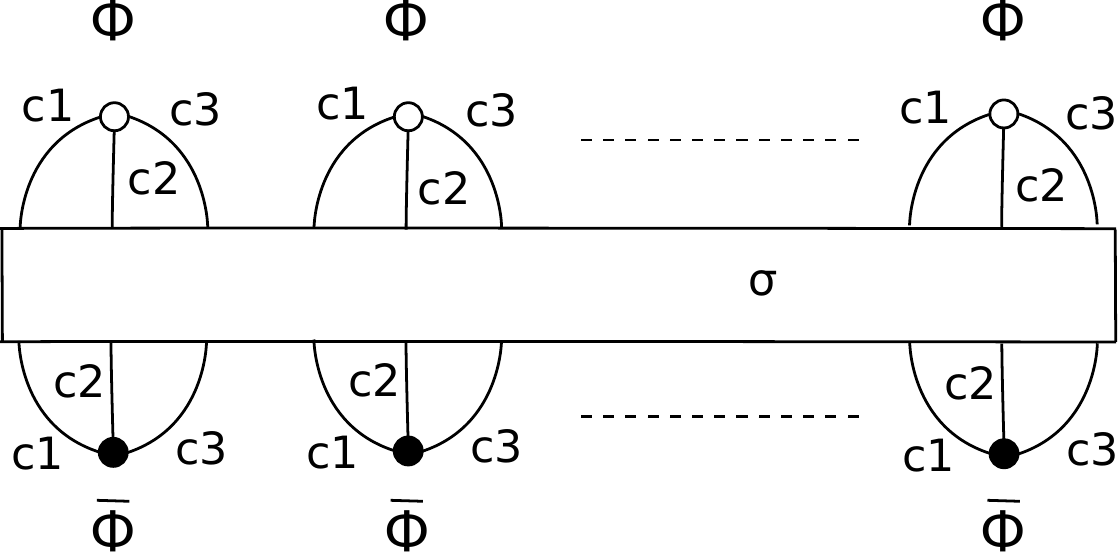}
\vspace{0.1cm}
\caption{ {\small Symmetric tensor contraction}} \label{fig:csym}
\end{minipage}
\end{center}
\end{figure}

The possible contractions between these fields can be drawn as the possible parings between two families of $n$ vertices with $d$ half-lines 
in the way given in  Figure \ref{fig:csym}. 

In other words, the determination of possible 
graph amounts to count the number of permutations 
\bea 
\s \in  S_{dn} \,,
\eea
permutations subject to the equivalence 
\bea 
\s \sim \gamma_1  \cdot \s \cdot \gamma_2 
\eea
where $ \gamma_i \in S_n \ltimes (S_d)^n =: \,S_n[S_d]$ (called the wreath product) act as follows. 
The $S_d^{\otimes  n } $  permutes independently the $d$-tuples
 of  indices for each of  the $n$ tensors (say $\Phi$); the $S_n$ acts 
by permuting the  $n$ tensors, equivalently it permutes the $n$ $d$-tuples among each other, while 
not changing their internal structure. 
The permutation  $\s$ acts pointwise on the full set of these $\{ 1 , \cdots , nd \} $  indices. 
If we write the $nd $ indices  on the $\Phi$'s as $i^{a}_{ \alpha} $ where 
$a $ runs from $1$ to $n$  and $\alpha $ runs from $ 1$  to $d$, with 
all indices with fixed $a$ attached to the same $\Phi$, the action of
 $ ( \gamma ; \gamma_1 , \cdots , \gamma_n ) \in  S_n[S_d]  $ with $\gamma \in S_n , \gamma_a \in S_d $ 
 for $a \in \{ 1 ,\cdots , d \} $ acts as 
 \bea 
 i^{a}_{ \alpha} \rightarrow i^{ \gamma ( a) }_{  \gamma_a ( \alpha ) } 
 \eea

Hence, the counting we are interested in is given by the number of classes in the double coset space
\be 
 S_n[S_d] \ses    S_{dn} \, / S_n[S_d]
\ee
Applying (\ref{hgh2}),  the counting can be recast as
\be
\cZ_{d;\, \sym} ( n ) =  { 1 \over {(n!)^2(d!)^{2n}}  } 
\sum_{C } (Z_C^{ S_n[S_d] \to S_{dn}})^2\; 
\Sym (C)
\label{ssd2}
\ee
In order to achieve this, we use similar generating function techniques
as developed in \cite{FeynCount}. We have the generating 
function of wreath products as 
\be
\cZ_{d}^{S_\infty[S_d]}(t, \vec x) = \sum_{n} t^n Z^{ S_n[S_d]}(\vec x)
= e^{  \sum_{i=1}^{\infty} \, \frac{t^i}{i}\,  
\Big[\sum_{q \vdash d} \prod_{\ell=1}^{d} 
\big(\frac{ x_{i\ell} }{ \ell } \big)^{\nu_\ell} 
\frac{1}{ \nu_\ell! } \Big] } 
\ee
where $\vec x = (x_1, x_2, \dots )$, and the partition $q= (\nu_\ell)_\ell$ of $d$
generate $\sum_{\ell} \ell \nu_\ell = d$.  Finally,
\bea 
\cZ_{d;\, \sym} ( n )  &=& 
\sum_{ p \vdash dn }   \Big(\hbox{Coefficient }  [  \cZ_{d}^{S_\infty[S_d]}(t, \vec x) , t^n x_1^{p_1} x_2^{p_2}\dots x_{dn}^{p_{dn}}  ]\Big)^2 \Sym (p)
\eea
A Mathematica program (see Mathematica code 6, in Appendix \ref{app:gap}) allows to obtain the sequences: 
For $d=2$ (matrix model fully symmetric invariants), for $n=1,\dots,13,$
\be
1,2,3,5,7,11,15,22,30,42,56,77, 101
\label{d213}
\ee
which gives, according to the OEIS, simply the number
of partition up to order of starting at $n=1$ up to order $13$. 
At order $n\geq 14$, the evaluation becomes challenging. 
This sequence might be coincide several known by OEIS 
(for instance A000041, A046054, etc). 

For $d=3$, one gets for $n=1,\dots,8$
\be
1,2,5,12,31,103,383,1731
\label{d38}
\ee
Last for $d=4$, we obtain for $n=1,\dots, 8,$
\be
1,3,9,43, 264, 2804, 44524, 1012456
\label{d47}
\ee
Both of \eqref{d38} and \eqref{d47} are new sequences according to 
the OEIS website.

\section{ Correlators  of tensor observables }\label{sec:correlators}  

We have already motivated the enumeration of the tensor 
model invariants in terms of classifying the possible interaction terms 
that can be added to the Gaussian term.  
Other perspectives suggest that there should be additional algebraic structures on these tensor invariants. 
In the context of matrix models, the string duals lead one to consider 
a state space with basis corresponding to the traces of matrices \cite{WittenTopPhase}.  
On this state space, there can be an interesting non-degenerate  pairing or  inner product. 
The pairings are related to correlators involving insertions of 
two of these general observables in the path integral. 
A ring structure  is also  a fruitful object of study containing information about the 
dual geometry \cite{WittenGroundRing}. 
 With this in mind, we can define a vector space with basis labelled by 
the tensor model invariants and study correlators involving insertions of two 
or more of the general invariants.   We write some formulae for 
correlators with two insertions of the observables we have classified in a Gaussian 
integral for colored tensors. We obtain 
some formulae in terms of permutation groups, with structure similar to the delta 
 function sums that appeared in the previous counting.  We will restrict attention to $d=3$.

Consider the Gaussian model 
\bea 
\cZ = \int d \Phi d \bar \Phi \; e^{ - { 1 \over 2 }  \Phi^{i_1 i_2 i_3 } \bar \Phi^{i_1 i_2 i_3 } } 
\eea
The index $i_a$ runs over $\{ 1 \cdots N_a \}$, for $ a \in \{ 1,2,3 \} $.  
The 2-point function is 
\bea 
\la \Phi^{ i_1 i_2 i_3    } \bar \Phi^{ j_1 j_2 j_3 } \ra = \delta^{ i_1 j_1 } \delta^{i_2 j_2 } \delta^{i_3 j_3 } 
\eea
The observables, invariant under $U(N) \times U(N) \times U(N)$, are labeled by permutations 
$ ( \s_1 , \s_2 , \s_3 ) $ subject to equivalence $ ( \s_1 , \s_2 , \s_3 ) \sim ( \gamma_1 \s_1 \gamma_2 , \gamma_1 \s_2 \gamma_2 , \gamma_1 \s_3 \gamma_2 ) $. We will write these observables as $ \cO_{ \sigma_1 , \sigma_2 , \sigma_3} $ with the 
understanding that 
\bea 
\cO_{ \s_1 , \s_2 , \s_3 } = \cO_{ \gamma_1 \s_1 \gamma_2 , \gamma_1 \s_2 \gamma_2 , \gamma_1 \s_3 \gamma_2 } 
\eea
The two-point function obtained by inserting in the above tensor  model integral the product of two such 
operators 
\bea 
\la \cO_{ \s_1 , \s_2 , \s_3 }  \bar \cO_{ \tau_1  , \tau_2 , \tau_3 } \ra =
{ 1 \over \cZ }  \int d \Phi d \bar \Phi \; e^{ - { 1 \over 2 }  \Phi^{ijk} \bar \Phi^{ijk} }  
 \cO_{ \s_1 , \s_2 , \s_3 }  \cO_{ \tau_1^{-1}   , \tau_2^{-1}  , \tau_3^{-1}  }
\eea
We consider the operators to be normal-ordered, i.e when we sum over Wick contractions, we 
do not include contractions between the $\Phi$'s within an operator $\cO_{ \s_1 , \s_2 , \s_3 } $
and $ \bar \Phi$'s within the same operator. The two-point function 
will be a function of $\s_i , \tau_i $ which is invariant when either the $\s_i$  or the $\tau_i$ are multiplied 
by $ \gamma_1 $ on the left and $\gamma_2 $ on the right.  
The answer is (see Appendix \ref{app:corrs} for the derivation)
\be
\label{corr2pt}  
 \la \cO_{ \s_1 , \s_2 , \s_3 } \bar  \cO_{ \tau_1  , \tau_2 , \tau_3 } \ra 
 = \sum_{ \mu_1 , \mu_2  \in S_n } 
 N_1^{ n } N_2^{ n }N_3^{ n }  \,\delta (\mu_1  \s_1 \mu_2 \tau_1^{-1}  \Omega_1 ) \delta (  \mu_1  \s_2 \mu_2 \tau_2^{-1}  \Omega_2 )  
\delta (  \mu_1  \s_3 \mu_2 \tau_3^{ -1}  \Omega_3 )   
\ee
Here $ \Omega_a = \sum_{ \sigma \in S_n } N_a^{ C_{ \sigma_a  } - n } \sigma $ and is in the group algebra of $S_n$. 
A transformation 
\bea 
&& \sigma_i \rightarrow \gamma_1 \sigma_i \gamma_2  \cr 
&& \tau_i \rightarrow \gamma'_1 \tau_i \gamma'_2 
\eea
can be absorbed by changing variables in the sums 
\bea 
&& \mu_1 \rightarrow \gamma_2'  \mu_1 \gamma_1 \cr 
&& \mu_2 \rightarrow \gamma_1' \mu_2 \gamma_2
\eea 
This shows that the correlator gives a 
pairing of the equivalence classes of permutation triples which we counted 
in Section \ref{sect:Counting}. 
Note that $\Omega_a $ commute with all permutations in $S_n$. 
 In the large $N_a $ limit, $\Omega_a \rightarrow 1 $. 
Then the 2-point correlator becomes an inner product which is diagonal 
on the equivalence classes, with positive diagonal values. This is an analog 
of the familiar { \it large $N$  factorization}  of matrix model, where different trace structures 
do not mix in the leading large $N$ limit. Here the two equivalence classes of invariants inserted (which are the analogs of 
trace structure for one-matrix invariants) have to be identical for a non-vanishing 
2-point correlator.  At subleading orders ${ 1 \over N_a} $,  different equivalence classes 
mixing under the inner product, with the mixing being controlled by the group 
multiplication in $S_n$.

The equation (\ref{corr2pt}) can be further simplified by defining 
$ \alpha_2 = \sigma_1^{-1} \sigma_2 ,  \alpha_3 = \sigma_1^{-1} \sigma_3 $ 
and $ \beta_2 = \tau_1^{-1} \tau_2 ,  \beta_3 =  \tau_1^{-1} \tau_3 $. 
\bea 
\langle \cO_{ \sigma_1 , \sigma_2 , \sigma_3 }  \bar{\cO}_{ \tau_1  , \tau_2 , \tau_3 } \ra
= n! \sum_{ \mu \in S_n } \delta ( \beta_2^{-1} \mu^{-1} \alpha_2 \mu \Omega_1 \Omega_2 ) 
                                       \delta ( \beta_3^{-1} \mu^{-1} \alpha_3 \mu \Omega_1 \Omega_3 ) 
\eea
This simplification is analogous to the one that happened in the counting delta functions 
(\ref{gam2})(\ref{TwoCoveringInterps}). 
The close parallels between counting and correlators exhibited 
by the permutation-TFT approach is a recurrent theme that has been encountered 
for example in \cite{countconst,quivcal,doubcos} in the context of AdS/CFT. 
The algebraic structures present in correlators, such as non-degenerate pairing 
(inner product) and product structure (related to insertion of three observables), 
are also of interest in the context of CFTs. We will comment 
on a 4D CFT context for studying the combinatorics and correlators of 3-index fields 
in Section \ref{sec:discussions}.

\section{ Summary and  discussion  }
\label{sec:discussions}

In this section, we will summarize the main results of this paper and outline extensions thereof. 
We then  discuss some conceptual questions raised by  the results of this paper, and describe associated technical investigations  that can be carried out. 

\subsection{ Summary  of main results } 

\begin{itemize} 

\item There is a counting of invariants made from $n$ copies of a
colored $d-$tensor, along with $n$ copies of the conjugate tensor given
in terms of a sum over partition of $n$  \eqref{ncopdtens}. This counting includes disconnected invariants (analogous to multi-traces
in matrix models). 
With this disconnected counting as input, the plethystic log function is used to 
generate the connected invariants.  Using these formuli, we generated the counting sequences to high order: \eqref{SumSymMu} and \eqref{ran4} give the disconnected counting for the rank 3 and rank 4 case,
respectively, whereas \eqref{r3conX} and \eqref{r4conX} give the connected counting for the rank 3 and 4, respectively.  

\item  We have shown that the counting of invariants of the $d-$tensors, with 
$n$ copies of $ \Phi $ and $n$ of $\bar \Phi$,  
 is equivalent to the counting of degree $n$  
 branched covers of the sphere with $d$ branch points (summed over the possible 
 genera of the covering space). Other geometrical interpetations in terms of covering spaces 
 are also discussed in Section \ref{sect:connect}. Permutation-TFTs, in conjunction  with the Burnside lemma from combinatorics and the links between fundamental groups, permutations and
 covering spaces given by algebraic topology, form  a unifying framework for exhibiting the different 
 geometrical interpretations. 

\item For the case $d=3$, the counting of tensor invariants is equivalent to the counting of 
 embedded  bi-partite graphs with $n$ edges
  and is  also related to the computation  of correlators of complex  matrix models. 

\item We studied a color-symmetrized counting, obtaining explicit formulae in terms of multi-variable 
generating  functions.  Key results are \eqref{s1s2s3}, \eqref{Eff-formulae}, \eqref{z4scdeltas} and \eqref{Eff-forms-d4}. 

\item The permutation techniques were used to give counting formulae for 
the tensor invariants in the cases of the more traditional non-colored tensor models. 

\item As a start towards investigating algebraic structures on the space of 
tensor observables provided by the Gaussian tensor model, we gave permutation 
group formulae for the 2-point correlator of the general invariants.
 We noted that the normal-ordered 2-point correlator 
gives an inner product, which is diagonalized by the equivalence classes of tensor invariants
(or, expressed another way, by the equivalence classes of branched covers of the 2-sphere)
 in the large N limit. This diagonality is a tensor model analog of large $N$ factorization of matrix models.

\end{itemize}

\subsection{Discussion} 

In this section, we  discuss some conceptual questions  raised by 
our results and  list some  related problems for investigation.

\subsubsection{ Braid orbits } 

Given a permutation triple, $ ( \tau_1 , \tau_2 , \tau_3) $. 
obeying 
\bea 
\tau_1 \tau_2 \tau_3 =1 
\eea
Color-symmetrization proceeds by group actions generated by $(C_1 , C_2)$ 
\bea 
&&  C_1 ( \tau_1 , \tau_2 , \tau_3 ) = ( \tau_2 , \tau_1 , \tau_1^{-1} \tau_2^{-1} ) \cr  
&& C_2 ( \tau_1 , \tau_2 , \tau_3 ) = ( \tau_1^{-1} , \tau_1^{-1} \tau_2 , \tau_2^{-1}  \tau_1^2 ) 
\eea
One checks that 
\bea 
C_1^2 = 1\,,  ~~ C_2^2 = 1\,, ~~ C_1 C_2 C_1 = C_2 C_1 C_2
\eea
This means that the group generated by $\{C_1,C_2\}$ contains
\bea 
\{ 1 , C_1 , C_2 , C_1C_2 , C_2 C_1 , C_1C_2C_1 \} 
\eea
and is $S_3$, the symmetric group of permutations of $3$ elements.  

Recall that this came from gauge-fixing $ ( \s_1 , \s_2 , \s_3 )$, using the gauge equivalence  in \eqref{gamma-equivs}
\bea 
( \s_1 , \s_2 , \s_3 ) \rightarrow ( 1 , \s_1^{-1} \s_2 , \s_1^{-1} \s_3 ) \equiv ( 1 , \tau_1 , \tau_2 ) 
\eea

There is another $S_3$ action on triples $ \tau_1 , \tau_2 , \tau_3 $ which multiply to $1$, which 
is generated by two braiding generators $B_1, B_2$ which act as follows 
\bea 
&& B_1 ( \tau_1 , \tau_2 , \tau_3 ) = ( \tau_2 , \tau_2^{-1} \tau_1 \tau_2 , \tau_3 ) \cr 
&& B_2 ( \tau_1 , \tau_2 , \tau_3 ) = ( \tau_1 , \tau_3, \tau_3^{-1} \tau_2 \tau_3 ) 
\eea
Again we have $ B_1^2 = B_2^2 =1 $ and $ B_1 B_2 B_1 = B_2 B_1B_2 $, so that   the group generated is $S_3$.

From the above description, there appear to be two similar but distinct 
$S_3$  actions -one coming from color-symmetrization and one from braiding. 
 Yet when we compute the number of braid orbits using 
Burnside's lemma, applying delta functions and  simplifying, we get the same answer as with 
color-symmetrized equivalence classes. Also computation with GAP gives the same counting. 
This  means that the formulae  \eqref{s1s2s3} and \eqref{Eff-formulae} give the counting of braid orbits. 
Braid orbits are of interest from the point of view of the topological classification of
polynomials  \cite{jozvo}.

It is natural to ask if the connection between color-symmetrized equivalence classes 
and braid orbits goes beyond the counting and holds for the actual orbits themselves. 
This would hold if a more direct connection between the two actions of 
$S_3$ on $ ( \tau_1 , \tau_2 , \tau_3 ) $ could be found, e.g by some appropriate 
change of variables.  Even at the level of counting, there is the question of whether 
the equality holds for $d$ higher than $3$. The cases $d =4,5$ should be a somewhat tedious 
but very doable problem.

\subsubsection{  Higher dimensional topology and low-dimensional covers} 

The primary motivations for the study of tensor models by physicists has been 
its connections to higher dimensional topology. With an improved understanding 
of counting problems associated with tensor models
 and with the aid of modern computational tools for group theoretic computations,
 one may ask if tensor models can provide a new perspective on counting problems 
  in topology studied in the mathematical literature  e.g. \cite{luostong,barnette}. 
For example, can we use tensor models to count triangulations of 3-sphere with specified 
numbers of vertices ? Another goal would be to try and extract information 
about continuum geometry from discrete computations, through mathematical 
connections such as that provided by the Riemann existence theorem -- we would need 
some form of higher dimensional generalizations of it. 

What is intriguing in the connection between tensor models and 
branched covers of the two-sphere we have developed here, is 
that it suggests that two dimensional holomorphic maps know 
about higher dimensional combinatoric topology. 
The study of dimer models - and the associated bi-partite graphs and Belyi maps - 
in connection with toric Calabi-Yau geometries is another example of 
physical links between  low-dimensional holomorphic maps and higher dimensional geometry
\cite{FHKVW,JRR,Yang,YangMac,SGS,ACD}.

\subsubsection{ Fourier transforms and finite $N$  effects  } 

In all the counting problems we have treated 
 in this paper,  we have treated $N$ -- the range of values taken 
 by the tensor index -- to be large. There are qualitative changes 
 in the  counting when $N$ is finite. For the case of matrices, this is 
 a consequence of Caley-Hamilton theorem which allows us to 
 write $ \tr (X^{N+1}) $ for an $N \times N$ matrix in terms of 
 products of lower traces. This has important implications in 
 string theory in the form the stringy exclusion principle 
 \cite{malstrom}. These finite $N$ effects have been studied in a variety 
 of multi-matrix systems \cite{BHR1,BCD1,KR1}. The key lesson is that 
 they are neatly characterized  by using permutations to describe
invariants (as we have done here) and then performing the  Fourier 
transform on permutation groups to go from ``permutations subject to constraints'' 
to appropriate representation theoretic data given by representations of permutation groups. 
The finite $N$  cutoffs are simple in terms of Young diagrams. The reason why representation 
theory of the permutation groups knows about the finite $N$ of $U(N)$  is Schur-Weyl duality. 
For an overview of how Schur-Weyl duality enters gauge-string duality see \cite{CMR1,SWrev,Gigravosc,doubcos}.

\subsubsection{ A gauge theory perspective on counting and correlators of tensor invariants } 

Consider a gauge theory, say  in 4 dimensions, with gauge group 
$U(N)^{\times 3}$. 
Choose the matter to be a Lorentz scalar which is complex and  transforms in the $ ( N , N , N ) $
of the gauge group. It is then a four dimensional field $ \Phi_{ i j k} ( x ) $. we may ask how to enumerate all the gauge invariant 
observables made from  $ \Phi_{ i j k} ( x ) $ in the large $N$  limit. The zero coupling limit 
is a conformal field theory, so we have an operator-state correspondence. 
The enumeration we gave in Section \ref{sect:Counting} is then counting physical (gauge-invariant) 
states that can be built from the scalar. The two-point correlators we computed give 
the CFT-inner-product on these states (for uses of operator-states corresponding in the 
context of AdS/CFT see for example \cite{WittenAdSCFT}).  
3-index fields have recently been of interest in the context of supersymmetric gauge theories 
 \cite{GaiottoMalda,tachyone}.

\subsubsection{Complex matrix models and 3-index tensor models : An intriguing  relation  } 

Consider a complex matrix model with Gaussian measure, with 
\bea 
\int dZ e^{ - { 1\over 2 } \tr ZZ^{\dagger} }  
\eea
where we have 
\bea 
 \la Z^{i}_j (Z^{\dagger})^k_l \ra = \delta^i_l \delta_j^k  
\eea
The holomorphic traces of $Z$ can be parametrized by permutations  $ \tau $ 
\bea 
\cO_{ \tau } ( Z ) = \sum_{ i_1 \cdots i_n =1 }^N 
 Z^{i_1}_{ i_{ \tau (1)  }} \cdots Z^{i_n}_{ i_{ \tau (n)  }}
\eea
subject to constraints 
\bea 
\cO_{ \tau } = \cO_{ \gamma \tau \gamma^{-1} } 
\eea
for $ \gamma \in S_n  $. This parametrization includes both single traces such as 
$ \tr (Z^3) $ and multi-traces such as $ [\tr (Z^2)](\tr Z)$.  The cycle structure 
of $\tau$ determines the numbers of single traces, double traces etc. 
Of particular interest in AdS/CFT are the correlators with one holomorphic and one anti-holomorphic observable. 
\be
{ |T_1| \over n! } { |T_2 | \over n! } \la \cO_{ \tau_1} ( Z )  \cO_{ \tau_2 } ( Z^{\dagger})  \ra 
\ee
The  natural normalization factors involve the sizes of 
the conjugacy classes corresponding $ \tau_1 , \tau_2 $ which have been denoted $T_1 , T_2$. 
It can be shown that the correlator  is a sum over triples of permutations \cite{cjr,Gal,brown} 
\be
 { |T_1| \over n! } { |T_2 | \over n! }  \la \cO_{ \tau_1} ( Z )  \cO_{ \tau_2 } ( Z^{\dagger})  \ra 
 = { 1  \over n!  }  \sum_{ \tau_1 \in T_1 } \sum_{ \tau_2 \in T_2} \sum_{ \tau_0 \in S_n } 
\delta ( \tau_1 \tau_2 \tau_0 ) N^{ C_{ \tau_0 }} 
\ee
 This shows that the correlator is a sum over branched covers of the 2-sphere, 
 branched over three points. The covers are summed with weight given by the  inverse  order 
 automorphism group  of the covers. This is a geometrical description of the Feynman 
 graphs (more bi-partite embedded graphs) of the matrix model. 
 
 In this paper, we have found  that observables of the 3-index tensor model 
 are parametrized by permutations $ \tau_1 , \tau_2 $ subject to conjugation equivalence
 (see equations \ref{gam2}  \ref{triplesequivs}). 
 These equivalence classes are precisely the Feynman graphs for the correlators of the complex matrix model described above.  
Feynman graphs of the matrix model become physical states (observables)  of
the  tensor model. 
As we saw the (normal-ordered) two-point correlators of  the tensor model provide an inner product 
on these observables. So in this case,  in a more than superficial sense, 
{\it  Feynman graphs of a matrix model 
have become states of a tensor  model}.  It would be interesting to 
unravel the proper interpretation and implications of this connection. 
 How general is it ? It has a flavor of being a dimensional uplift, which is often related to categorification
 (see further discussion of the connection between (refined) graph counting and three-dimensional 
 permutation-TFTs in \cite{refcount}).  
This should be better understood both from a physical and a mathematical point of view. 
Note that the usual physical argument  for tensor models being a higher dimensional generalization 
of matrix models relies on interpreting the indices as being dual to simplexes. Here we are seeing 
an extra dimension from the tensor model by considering counting and correlators of invariants, 
which are objects built after contracting away all the indices.

\section*{Acknowledgements}
 SR thanks the Perimeter Institute, Waterloo, Canada, for hospitality during part of his sabbatical year and for  providing an excellent environment for  research interactions. In  particular he thanks  his host Laurent Freidel for wide-ranging physics discussions. JBG thanks Pedro Viera for helpful discussions on Mathematica programming.
We also thank for discussions Robert de Mello Koch, David Garner, Brian Wecht and Congkao Wen.   SR is supported by STFC Grant ST/J000469/1, String Theory, Gauge Theory, and Duality. Research at Perimeter Institute is supported by the Government of Canada through Industry Canada and by the Province of Ontario through the Ministry of Research and Innovation.

\section*{ Appendix}

\appendix

\renewcommand{\theequation}{\Alph{section}.\arabic{equation}}
\setcounter{equation}{0}

\section{Orbit-stabilizer theorem, Burnside's lemma, size  of conjugacy classes}
\label{app-group-basics}

We gather, in this appendix, basic facts about conjugacy 
classes in the symmetric group $S_n$, the group of all $n!$ permutations of 
$n$ objects, and about finite groups acting on finite sets.  Further discussion 
of these topics can be found, for example  in \cite{Cameron}.

\begin{definition}[Cycle-type]
\label{def:cycl}
Two permutations are of the same cycle type or have the
same cycle structure if the unordered list of sizes of their
cycles coincide. 
\end{definition}

\noindent {\bf Example}: Consider $\sigma_1$ a permutation defined 
by its cycles $(123)(4)(5)(67)$
the list of the sizes of the cycles of $\sigma_1$ is $(3,1,1,2)$. 
Note that the order in which appear 3,1,1,2 is not relevant. 
Consider another permutation $\sigma_2$
such that $(12)(3)(456)(7)$, then  $\sigma_1$ and $\sigma_2$ have 
the same cycle type. 

The cycle type of a permutation in $S_n$ determines a list $p = ( p_1 , \cdots  , p_n ) $
of numbers $p_i$ of cycles of length $i$. The list $p$ is a partition of $n$ : 
\bea 
n = \sum_i i p_i 
\eea

\begin{proposition}[Conjugacy class]\label{prop:cc}
Two permutations $ \alpha $ and $\alpha'$  have the same cycle type if and only 
if they are related to each other by conjugation, i.e $\alpha' = \sigma \alpha  \sigma^{-1} $ for  
some $\sigma $. 

\end{proposition}
\proof 
($\Leftarrow$) The action under conjugation preserves the cycle structure. 
Indeed, consider $\alpha$ and $\sigma$ two permutations. From $\sigma \alpha \sigma^{-1}(\sigma(x))=\sigma(\alpha (x))$,
one has for any cycle of a given permutation
\be\label{conjsubs} 
\sigma (a_1,\dots, a_2) \sigma^{-1} =  (\sigma(a_1),\dots, \sigma(a_2))
\ee
($\Rightarrow$) Consider $\sigma_1$ and $\sigma_2$ with
same cycle type. Construct first a bijection $\phi$ between the cycles
of these permutations mapping cycles with the same size
one onto another ($\phi$ may be not unique). For a pair of cycles $s_1=(a_1,\dots, a_q)$ of
$\sigma_1$ 
and $s_2=(b_1,\dots, b_q)$  of $\sigma_2$ linked by $\phi$, 
namely $\phi(s_1) =s_2$, 
construct a bijection $\sigma$ such that  $\sigma (a_i)=b_i$
($\sigma$ may be not unique as well).
Then one checks that $\sigma s_1 \sigma^{-1} = s_2$ 
and that $\sigma \sigma_1 \sigma^{-1} =\sigma_2$. 

\qed

\noindent 
{ \bf Burnside's lemma } \\

Consider a finite set $X$ and a finite group $G$ acting on $X$. 
Consider $x\in X$ and the application $F_x: G \to X$
such that $g \mapsto gx$. Note that the image of $F_x$, 
$\Im (F_x)  = G x$ is the orbit of $x$ in $X$ whereas 
the kernel $\ker (F_x) = G_x$ of $F_x$ is the stabilizer of $x$
in $G$. The {\it orbit-stabilizer theorem} states that the size of the 
orbit generated by the group action on an element $x$ is the ratio
of the group size divided by the size of the subgroup which leaves the 
element $x$ fixed. In equations 
\be
|G x| = [G: G_x] = \frac{|G|}{|G_x|}
\label{orbit}
\ee
The following statement holds. 
\begin{proposition}[Burnside's lemma]\label{prop:BL}
The number of orbits of the $G$ -action on $X$, denoted
$ |X / G |$  is given by average number of fixed points of the 
group action. More explicitly,  
\be
|X / G | = \frac{1}{|G|}\sum_{g \in G} X^g
\ee
where $X^g= \{x \in X, \, gx=x\}$ is the set of 
fixed point of $g$. 
\end{proposition}
\proof Let us observe that 
\be
\sum_{x \in X}|G_x|  = \sum_{x \in X}[ \sum_{g\in G/ gx=x} 1]
 = \sum_{g \in G}[ \sum_{x\in X/ gx=x} 1]  = \sum_{g \in G} X^g
\ee
Then inverting the relation \eqref{orbit}  and summing over $x$ yields
\bea
 \sum_{g \in G} X^g &=& \sum_{x \in X}|G_x| 
=|G| \sum_{x \in X}\frac{1}{|Gx|} 
 = |G| \sum_{x \in X}\sum_{A \in X/G/ \,x\in A} \frac{1}{|A|} 
 = |G| \sum_{A \in X/G}\sum_{x\in A} \frac{1}{|A|} \crcr
&=& |G| \,|X/G| 
\eea
where we used the fact that the classes in $X/G$ determine
a partition of $X$. 

\qed

If one is interested in the  number of elements in a conjugacy class
of symmetric group, then, by Proposition \ref{prop:cc}, it is enough to look at their unique cycle type.  Precisely, the following statement holds. 

\begin{proposition}[Size of conjugacy classes]\label{prop:ncc}
Consider the conjugacy class $T_p $ in the symmetric group $G=S_n$ with cycle type entirely determined by the list $ p =( p_1, p_2,\dots, p_n)$, where $p_i$ gives the number of cycles of size $i$.
This list forms a partition of $n$ since $n = \sum_i i p_i $. 
 Then the size of the conjugacy class $|T_p|$  is given by 
\be
|T_p| = \frac{n!}{ \Sym~  p  }\,, \qquad  \Sym ~ p  = \prod_{i=1}^n (i^{p_i}) ( p _i !)
\ee
where $\Sym ~ p $  is the number of elements of $S_n$ commuting with any permutation 
in the conjugacy class $T_p$. 
\end{proposition} 
This is an application of the orbit stabilizer theorem, for the case where the group $S_n$ acts on itself by conjugation. 

$\Sym ~  p $  can be computed as follows. For $\alpha \in T_p$, we are looking 
for $\sigma \in S_n $ such that $ \sigma \alpha \sigma^{-1} = \alpha $. 
As we saw in \eqref{conjsubs}, conjugating $\alpha$ with $ \sigma$  
amounts to replacing the integers $j $ in the cycles of $\alpha $ by $ \sigma ( j ) $. 
This $\sigma $ transformation of the cycles of $ \alpha $ can leave the cycles fixed, 
or exchange cycles of the same length. Focusing on cycles of length $i$, 
of which there are $p_i$, $\sigma$  can cycle the numbers within a cycle. 
For a cycle of length $i$ there are $i$ of these cyclic permutations. 
So there are $i^{p_i}$ cyclic permutations $ \sigma$ which just cycle 
the integers within cycles of length $i$ in $\alpha$, thus leaving $\alpha $ unchanged. 
Then, there are permutations of exchanging the $p_i$ different cycles.
In all, we get  $\prod_{i=1}^n (i^{p_i}) ( p _i !)$ as stated above.

\section{ Symmetric group delta functions to generating functions for counting } 
\label{app:s2n}
In this appendix, we address the evaluation of formal 
sums appearing as $S^{(3)}_{[2,1]} $ and $S^{(3)}_{[3]}(n)$ in \eqref{s1s2s3}
and $S^{(4)}_{[2,1^2]}$, $S^{(4)}_{[3,1]}$, $S^{(4)}_{[2^2]}$ and 
$S^{(4)}_{[4]}$ appearing
in \eqref{s1ps2p}.

Let us start by $S^{(3)}_{[2,1]} $ and find a way to perform this sum.  We have 
\be
 S^{(3)}_{[2,1]}= \sum_{ \gamma , \sigma \in S_n} \delta ( \gamma^2 \sigma \gamma^{-2} \sigma^{-1} ) 
\ee
For every  partition $p$ of $n$, 
$n = p_1 + 2 p_2 + \cdots\,, $
there is a permutation $\sigma$ of cycle of type $p$, i.e., $\s$ has 
$p_1$ cycles of length $1$, $p_2$ cycles of length $2$, etc. 
Let us denote this by $\s \in p$. 
Let $T_p$ be the sum of permutations in the cycle-type $p$ in the group 
algebra $\IC(S_n)$:
\be 
T_p = \sum_{ \sigma \in p } \sigma
\ee 
Consider the sum, still with value in $\IC(S_n)$,  
\be 
Z^{(2)} ( n ) = \sum_{ \gamma \in S_n } \gamma^{2}
=  \sum_{ p \vdash n }  Z^{(2) }_p { T_p \over |T_p| }  
\ee
The sum of $\gamma^2$ commutes with each element of $S_n$ ($\forall \s\in S_n$, $\sum_{\gamma} \s\gamma^2 \s^{-1} =\sum_{\gamma} (\s\gamma\s^{-1})^2  =\sum_{\gamma} \gamma^2$), so it is a sum over 
complete conjugacy classes $T_p$, each which some weight. 
We have defined  ${ Z^{(2) }_p \over  |T_p| } $ to be the coefficient of $ T_p $ in the sum of $\gamma^2$, 
where $|T_p|$  is the number of permutations in the conjugacy class 
corresponding to cycle-type given by $p$ (see Proposition \ref{prop:cc}).
Similarly, we can define 
\be 
\sum_{ \gamma \in S_n } \gamma  = \sum_{ p \vdash n  }  Z^{(1) }_p { T_p \over |T_p| } 
\ee
In this case,  
\be 
Z^{(1)}_p = |T_p| = {n!  \over { \prod_i i^{p_i} p_i!} } 
= { n! \over \Sym ( p )  }
\ee
Now there is a generating function for $Z^{(1)}_p $ given by 
\be
Z^{(1)} ( t ,\vec x)=Z^{(1)} ( t , x_1 , x_2 , \cdots  )  = e^{\sum_{ i=0}^{\infty }{  t^i x_i \over i}  } = \sum_{ n=0}^{ \infty } 
{ t^n \over n! } 
 \sum_{ p \vdash n } Z^{(1)}_{p} \prod_{i} x_i^{p_i}
\ee
where $\vec x=(x_1,x_2,\dots)$. 
When we square a permutation, all odd cycles become odd cycles again, 
whereas all even cycles split in two of half the length of the formers.
As a result the generating function for $Z^{(2)}_p $ is 
\be
Z^{(2)} ( t , \vec x)   = Z^{(1)} ( t , x_1, x_2 = x_1^2 , x_3 , x_4 = x_2^2 , \cdots )
  = \sum_{ n=0}^{ \infty }  { t^n \over n! }  \sum_{ p \vdash n } Z^{(2)}_{p} \prod_{i} x_i^{p_i} 
\ee
We can finally write 
\bea\label{S2nKeyFormula}  
S^{(3)}_{[2,1]} &=& { 1 \over n! }  \sum_{ p \vdash n  } Z^{(2)}_p  \, \Sym (  p )    \cr 
 &=& 
\sum_{ p \vdash n }    \hbox{ Coefficient }  [  Z^{(2)} (  t , \vec x) , t^n x_1^{p_1} x_2^{p_2}\dots x_n^{p_n}  ] \times\big[  \prod_{i=1}^n i^{p_i} p_i!  \big]  
\eea
where, given a partition $p$ of $n$, $ \hbox{ Coefficient }  [  Z^{(2)} (  t , \vec x) , t^n x_1^{p_1} x_2^{p_2}\dots x_n^{p_n}  ] $ is  the coefficient 
of the monomial $t^n x_1^{p_1} x_2^{p_2}\dots$ $x_n^{p_n} $ in the series $Z^{(2)}$. This is easily programmable in Mathematica
(see Mathematica code 2, in Appendix \ref{app:gap}) and 
one gets (from $n=1$ to $n=13$)
\be
1,2,5,17,59,265,1095,6342,33966,219968,1333654,9930505,70419371,\dots
\ee 

Another quantity which appears in the computation of the invariants constructed from 3-index invariants is  
\be
 S^{(3)}_{[3]} ={ 1 \over n! } \sum_{ \gamma , \sigma \in S_n } \delta  ( \gamma^3 \sigma^3  ) 
\ee
Consider the following element of the group algebra of $S_n$ : 
\be 
Z^{(3)} ( n ) = \sum_{ \gamma \in S_n } \gamma^{3}
=     \sum_{ p \vdash n }  
Z^{(3) }_p { T_p \over |T_p| }  
\ee
In terms of these 
\be
 S^{(3)}_{[3]}  = { 1 \over n! } \sum_{ p \vdash n } \left  (  {  Z^{(3) }_p \over |T_p| }  \right )^2  |T_p| 
              ={ 1 \over (n!)^2  }  \sum_{ p \vdash n }  \left (   Z^{(3) }_p  \right  )^2 \Sym ( p ) 
\ee
In an analogous way than before, there is a generating function 
\be
Z^{(3)} ( t , \vec x )  = \sum_{ n=0}^{\infty }  t^n \sum_{ p \vdash n  } { Z^{(3)}_p \over n! }  \prod_{ i=1}^n x_i^{p_i}  
\ee
When we take the cube of a permutation, any cycle of length divisible by $3$ becomes a triple of $1$-cycles. Any other cycle stays a cycle of the same length. Hence, one has
\be\label{Z3genfun} 
Z^{(3)} ( t , x_1 , x_2 , \cdots )  = Z^{(1)} ( t , x_i \,\vert_{  x_i   \rightarrow { (x_{i/3})^3} \hbox{\tiny{  iff $ i $ is divisible by 3 } } } )
\ee 
So, we have 
\be\label{S3nKeyFormula} 
 S^{(3)}_{[3]} = \sum_{ p \vdash n }  (\hbox{Coefficient } [  Z^{(3) } ( t , \vec x )  , t^n \prod_i x_i^{p_i} ]   )^2 \times \big[\prod_i i^{p_i } p_i! \big]
\ee
This is easily calculable in Mathematica (see Mathematica code 2 in Appendix \ref{app:gap}) and we list the numbers 
starting at $n=1$  up to $n=13$ as 
\bea 
1,1, 2, 4, 5, 13,29 , 48 , 114 , 301 , 579 , 1462 , 4198,\dots 
\eea
The first few terms can be easily checked  in GAP by directly summing pairs of permutations 
subject to $ \gamma^3 \sigma^3 ={\rm id }$.

 Counting the case of tensors with 4-indices, say $Z_{4;\,\col}(n)$, we encounter the sum $ S^{(4)}_{[3,1]} ( n) $ which is similar to that $ S^{(3)}_{[2,1]} ( n)$, but $Z^{(2)}$ is replaced with $Z^{(3)}$:  
\bea 
 S^{(4)}_{[3,1]}(n) &=& { 1 \over n! }  \sum_{ p \vdash n  } Z^{(3)}_p  \, \Sym (  p )    \cr 
 &=& 
\sum_{ p \vdash n }    \hbox{ Coefficient }  [  Z^{(3)} (  t , \vec x ) , t^n x_1^{p_1} x_2^{p_2}\dots x_n^{p_n}  ] \times \big[  \prod_{i=1}^n i^{p_i} p_i!  \big]   \label{s431}
\eea
where $Z^{(3)} $ is as given above in (\ref{Z3genfun}). 
Using still Mathematica, this can be programmed and we 
get (see Appendix \ref{app:gap}, Mathematica code 2), for
$n=1$ to $n=13$, 
\be
1,2,4,12,27,103,391,1383,6260,32704,149045,812696,5034682\dots
\ee

Still in the rank 4 case, one finds the sum 
 \bea 
 S^{(4)}_{[4]} ( n ) = { 1 \over n! } \sum_{ \s , \gamma  \in S_n } \delta ( \gamma^4 \sigma^4 ) 
 \eea
Following the above arguments, we will define 
$Z^{(4) } ( t , \vec x ) $ by substituting in $ Z^{(1)} ( t , \vec x ) $ 
\bea 
x_i &&  \rightarrow (x_{i/4})^4   \hbox{ for  $i = 4q $ with integer $q$ } \cr 
      && \rightarrow (x_{i/2})^2 \hbox{ for $i = 4q+2 $ } \cr 
       && \rightarrow x_i \hbox{ for $i =  4q+1 $ or $i =   4q+3  $ }   
\eea 
Then 
\be\label{s44}
 S^{(4)}_{[4]} ( n ) =  \sum_{ p \vdash n }  (\hbox{Coefficient} [Z^{(4)} ( t , \vec x )  , \prod_i x_i^{p_i} ]   )^2 \times \big[\prod_i i^{p_i } p_i!  \big] 
\ee
Some  terms of this sequence, starting from $n=1$ up to $n=13$  
 (see Mathematica code 2 in Appendix \ref{app:gap})
\bea 
1, 2, 3,11,27,93 , 233, 978 , 3156, 13280, 44476 , 205 611 , 796091  \dots 
\eea
The first few terms are quickly checked by directly summing the delta function over the symmetric 
group with  GAP, but this soon becomes prohibitive, and the generating function method is much more efficient. 

Counting rank 4 tensor invariants up to color permutation leads to 
another sum given by
\bea 
 S^{(4)}_{[2,1^2]}( n ) = { 1 \over n! } \sum_{ \gamma , \s_1 , \s_2 \in S_n } \delta ( \gamma \s_1 \gamma^{-1} \s_1^{-1} ) 
 \delta ( \gamma^2 \s_2 \gamma^{-2} \s_2^{-1} ) 
\eea
For $ \gamma $ in a conjugacy class  given by $p$, 
let us define $ Sq ( p ) $ to be the cycle structure of $\gamma^2 $:
\bea 
&& (Sq ( p ) )_{ 2j} =   2 p_{4j} \cr 
&& (Sq ( p ) )_{ 2j +1 } =    p_{2j+1 }  + 2 p_{4j+2 } 
\eea
As $ \gamma $ runs over all the partitions $p$, we have 
\bea 
S^{(4)}_{[2,1^2]} && =  { 1 \over n! } \sum_{ p \vdash n }  |T_p|   ~ \Sym ( p ) ~ \Sym ( Sq (p) ) \cr 
                && =  \sum_{ p \vdash n } \Sym ( Sq (p) )\cr 
                 && =  \sum_{ p \vdash n } \prod_{ j =1 }^{ \lfloor { n \over 2 } \rfloor } (2j)^{2  p_{4j} } ( 2 p_{4j} )! \prod_{ j=0}^{ \lfloor { n \over 2 } \rfloor } ( 2j+1)^{ p_{2j+1} + 2 p_{ 4j+2} }   ( p_{2j+1} + 2 p_{ 4j+2} ) ! 
\label{s4212}
\eea
This is calculated with Mathematica for $n$ in the range $ 1$ to $10$
(see Mathematica code 3 in Appendix \ref{app:gap}) as: 
\bea 
1,4,15, 83, 385 , 2989 , 20559 , 203992 , 1827640, 21864590  \dots 
\eea
The first few terms are checked against GAP which calculates the delta functions directly. 

Finally, one can transform $S^{(4)}_{[2^2]}(n)$ in the following way:
\bea \label{s422}
S^{(4)}_{[2^2]} (n ) &=& { 1 \over  n! } \sum_{ \gamma \in S_n} \sum_{\s_i \in S_n}
\delta(\s_1^2\gamma^2\,) \delta(\gamma^{2} \s_2 \gamma^{-2}\,\s_{2}^{-1} ) \cr 
&=& { 1 \over  n! } \sum_{ \gamma , \alpha  \in S_n} \sum_{\s_i \in S_n}
\delta(\s_1^2 \alpha ) \delta ( \alpha^{-1}  \gamma^2 )  \delta(\alpha   \s_2 \alpha^{-1} \s_{2}^{-1} ) \cr 
&=&  { 1 \over  n! } \sum_{ \alpha \in S_n } \delta ( \alpha Z^{(2)} ( n ) ) 
 \delta ( \alpha^{-1}  Z^{(2)} ( n )  )\,  \Sym ( \alpha ) \cr 
& =& { 1 \over n ! } \sum_{ p \vdash n } \sum_{ \alpha  \in [p] }  
 \delta ( \alpha Z^{(2)} ( n ) ) \delta ( \alpha^{-1}  Z^{(2)} ( n ) ) \,\Sym ( \alpha ) \cr 
 &=&  { 1 \over n ! } \sum_{ p \vdash n  }     \sum_{ \alpha \in p } (\frac{ Z^{(2)}_p}{|T_p|} )^2\, \Sym ( p  ) \cr 
 & =&   { 1 \over n ! } \sum_{ p \vdash n  } { n! \over \Sym ( p ) }  (  Z^{(2)}_p )^2\, \frac{1}{|T_p|^2}\Sym ( p  )\cr 
 & =& { 1 \over (n !)^2 } \sum_{ p \vdash n  }  (  Z^{(2)}_p )^2 (\Sym(p))^2 \cr 
 & =&  \sum_{ p \vdash n }     \Big( \hbox{Coefficient}[  Z^{(2)} (  t , \vec x ) , t^n x_1^{p_1} x_2^{p_2}\dots x_n^{p_n}  ]  \, \Sym(p)\Big)^2 
\eea
Doing this with Mathematica (see Appendix \ref{app:gap} code ), we get for $n$ from $1$ to $12$. 
\be
1,4,17,105,685,5825,54013,585018,6873522,90254150,1275023778,
19651966895
\ee
The first few agree with GAP.

\section{Derivations for  correlator computations }\label{app:corrs} 

\begin{figure}[h]\begin{center}
\begin{minipage}[t]{.8\textwidth}\centering
\includegraphics[angle=0, width=7cm, height=2cm]{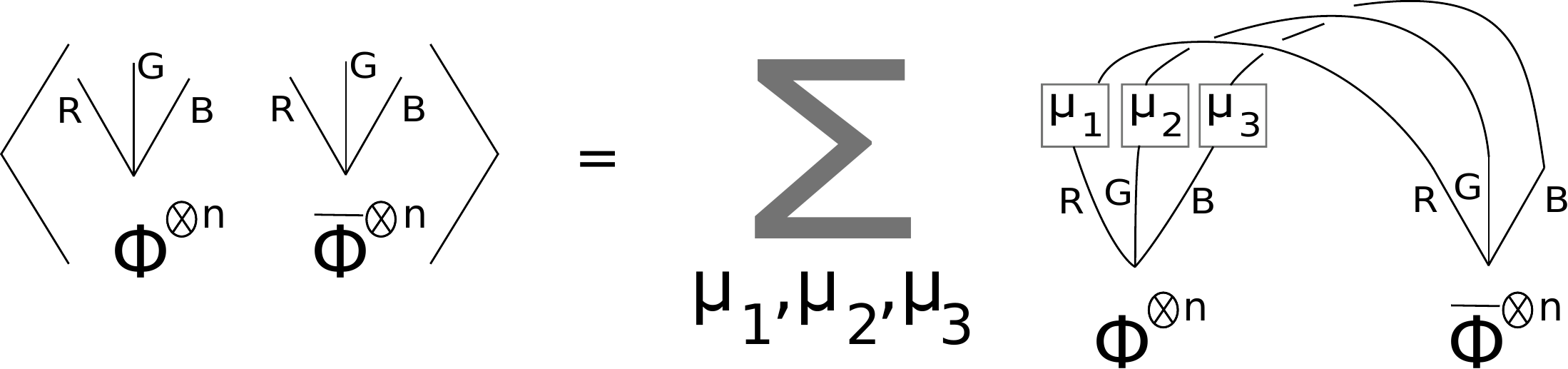}
\vspace{0.1cm}
\caption{ {\small Basic correlator in a diagrammatic form}} \label{fig:bascorr}
\end{minipage}
\end{center}
\end{figure}

In this section, we explain the derivations of the formulae for correlators 
in terms of delta functions over symmetric groups, which are best expressed 
in diagrammatic form. For some CFT applications of such techniques for correlators 
see \cite{cr,KR1,BHR1,BCD1,quivcal}.  
When we use the basic 2-point correlator and apply Wick's theorem to 
calculate the correlator of $n$ copies of $\Phi $ with $n$ copies of $\bar \Phi$, we get 
a sum over Wick contractions. This is a sum over permutations
which expresses as  
\bea 
&& \la \Phi^{i_1 , j_1 , k_1 } \cdots \Phi^{i_n , j_n , k_n }   ~ 
\bar \Phi^{i_1 , j_1 , k_1 } \cdots \bar  \Phi^{i_n , j_n , k_n } \ra \cr 
&& 
= \sum_{ \mu_1 , \mu_2 , \mu_3 \in S_n } \delta^{ i_1 , i_{ \mu_1 (1)} } \cdots  \delta^{ i_n  , i_{ \mu_1 (1)} } ~~ 
\delta^{ j_1 , j_{ \mu_2 (1)} } \cdots  \delta^{ j_n  , j_{ \mu_2 (n)} }  ~~ 
\delta^{ k_1 , k_{ \mu_3 (1)} } \cdots  \delta^{ k_n  , k_{ \mu_3  (n )} } 
\eea
It is convenient to describe this diagrammatically as in Figure \ref{fig:bascorr}.

\begin{figure}[h]\begin{center}
\begin{minipage}[t]{.8\textwidth}\centering
\includegraphics[angle=0, width=2cm, height=2.3cm]{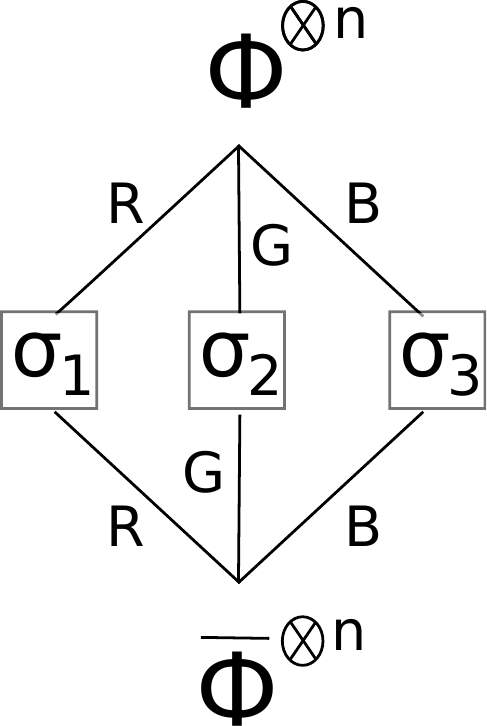}
\vspace{0.1cm}
\caption{ {\small Observable as a diagram}} 
\label{fig:Obsdiag}
\end{minipage}
\end{center}
\end{figure}

Let us also draw the observable parameterized by $\s_1 , \s_2 , \s_3 $ 
using a similar diagram of Figure \ref{fig:Obsdiag}. This is simplified version of the diagram in Figure \ref{fig:sss0}.

\begin{figure}[h]\begin{center}
\begin{minipage}[t]{.8\textwidth}\centering
\includegraphics[angle=0, width=12cm, height=3cm]{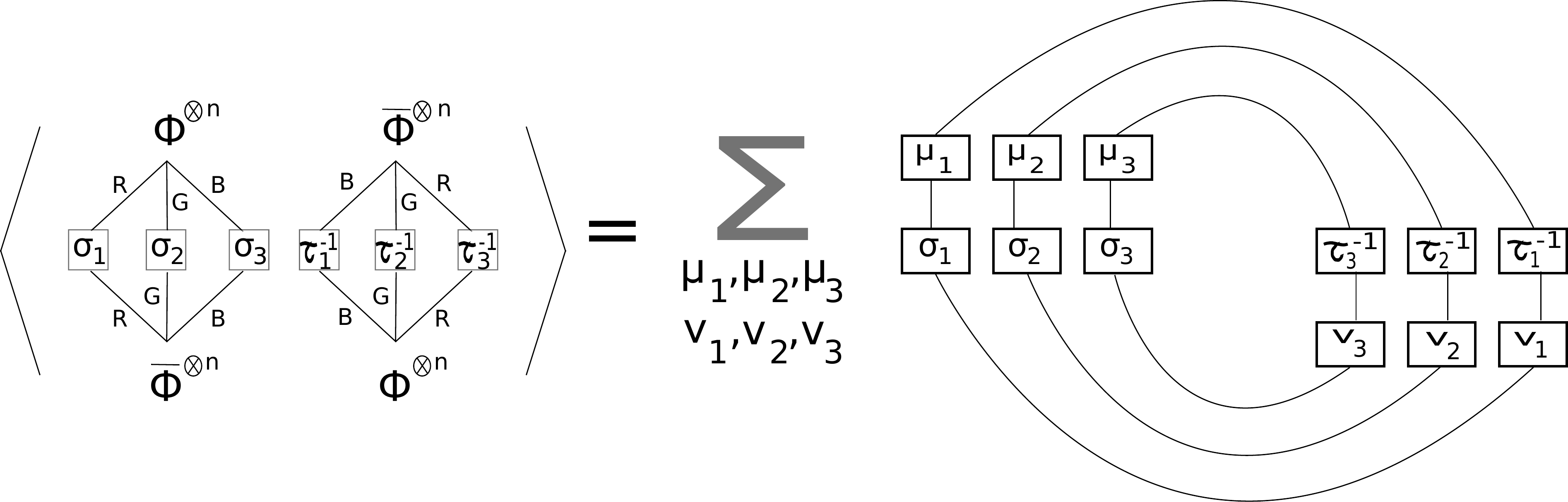}
\vspace{0.1cm}
\caption{ {\small Two-point function as a diagram}} \label{fig:diag2pt}
\end{minipage}
\put(-330,95){$\Big\la\cO_{\s_1,\s_2,\s_3}
\bar{\cO}_{\tau_1,\tau_2,\tau_3}\Big\ra=$}
\end{center}
\end{figure}

Similarly, draw the two-point function in a diagrammatic form 
given in Figure \ref{fig:diag2pt} and use the 
diagrammatic expression of the Wick contractions (Figure \ref{fig:bascorr})
in this correlator. As stated before, we are taking the observables to be ``normal ordered''
so we only allow contractions to take place between the $\Phi$'s from the first observable to the $\bar \Phi$'s 
from the second (parametrized by $\mu_a$) and  between the $\bar \Phi$'s from the first observable to the $ \Phi$'s 
from the second (parametrized by $\nu_a$).

The final step is a simple diagrammatic straightening, to recognize that the correlator is 
a product of three traces of sequences of permutations 
\be
\la \cO_{ \s_1 , \s_2 , \s_3 } \bar{\cO}_{ \tau_1 , \tau_2 , \tau_3 } \ra 
= \sum_{\mu_i \in S_n  }  \sum_{\nu_i \in S_n  }
\tr_{ V_1^{ \otimes n } } ( \s_1 \mu_1  \tau^{-1}_1 \nu_1 ) \tr_{ V_2^{ \otimes n } } ( \s_2 \mu_2  \tau^{-1}_2 \nu_2  )
\tr_{ V_3^{ \otimes n } } ( \s_3 \mu_3 \tau^{-1}_3   \nu_3 ) 
\ee
Now if $V$ is an $N$-dimensional space with basis $e_{i} $ for $i=1 \cdots N$. 
We have 
\bea 
 \tr_{ V^{ \otimes n } } ( \sigma ) 
& = &   < e^{i_1}\otimes  \cdots \otimes e^{i_n } | \sigma | e_{i_1} \otimes e_{i_n} > =  < e^{i_1}\otimes  \cdots \otimes e^{i_n }  | e_{i_{ \sigma(1) }}  \otimes e_{i_{ \sigma(n)} } > \cr 
& = & \delta^{ i_1}_{ i_{ \sigma(1)} } \cdots  \delta^{ i_n}_{ i_{ \sigma(n)} }= N^{ C_{ \sigma} } 
\eea
The repeated $i$ indices are summed since we  are taking a trace. 
$C_{ \sigma }$ is the number of cycles in the permutation $\sigma$. 
It is instructive to see how the last step works in a simple example, where $n=2$. 
If $ \sigma = (1)(2)$ is the identity permutation, then 
\be
  \delta^{ i_1}_{ i_{ \sigma(1)} }  \delta^{ i_2}_{ i_{ \sigma(2)} }
 =  \delta^{ i_1}_{ i_{ 1} }  \delta^{ i_2}_{ i_{ 2} } 
 = N^2   
\ee
If $ \sigma = (12)$ is the swop, we have instead:
\be   \delta^{ i_1}_{ i_{ \sigma(1)} }  \delta^{ i_2}_{ i_{ \sigma(2)} }
 =  \delta^{ i_1}_{ i_{ 2} }  \delta^{ i_2}_{ i_{ 1} } 
 = \delta^{i_1}_{i_1}    = N 
\ee
Thus, we see that the power of $N$ is the number of cycles in the permutation.

\begin{figure}[h]\begin{center}
\begin{minipage}[t]{.8\textwidth}\centering
\includegraphics[angle=0, width=2.5cm, height=2.5cm]{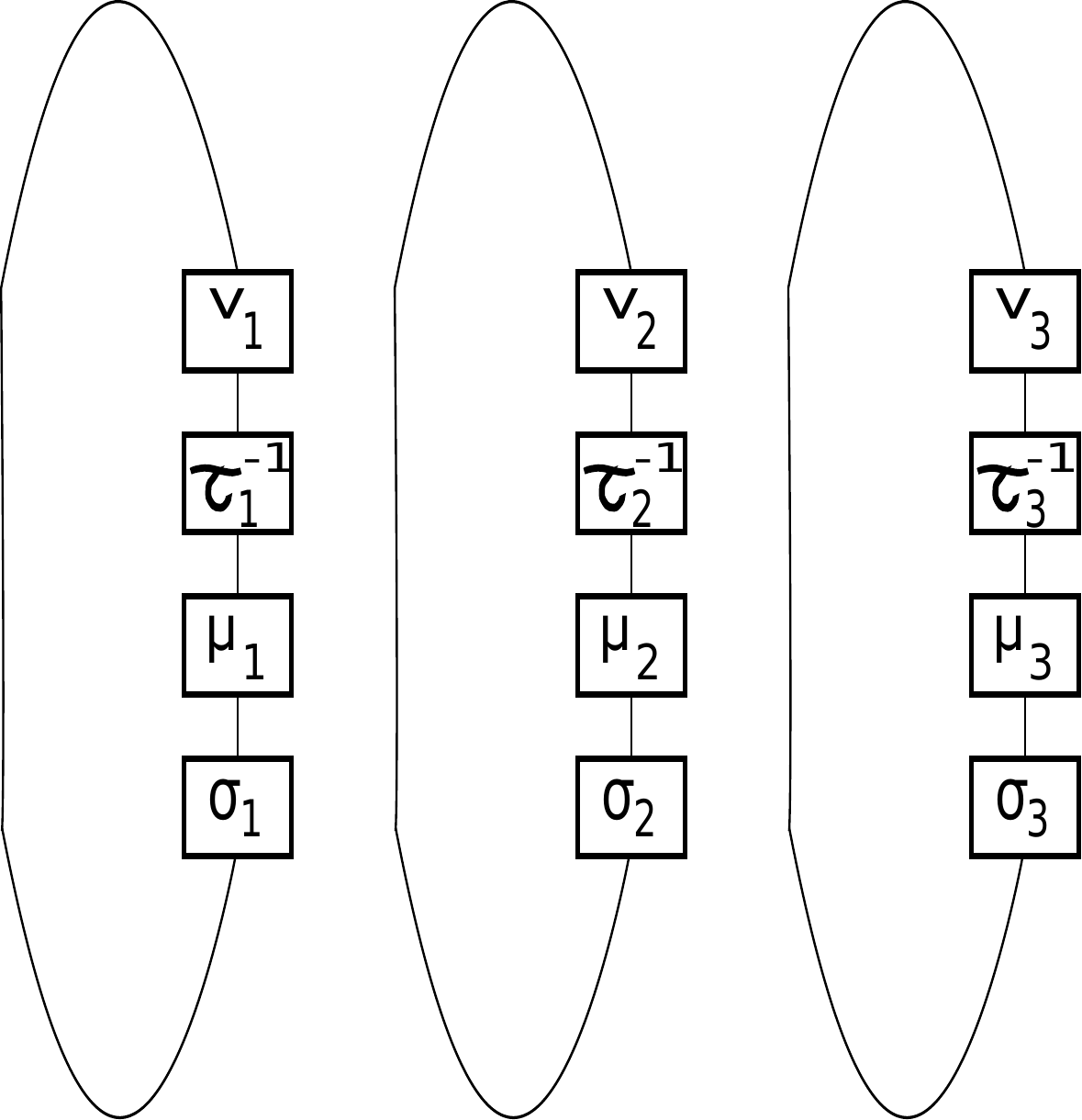}
\vspace{0.1cm}
\caption{ {\small Straightening the traces}} \label{fig:strTr}
\end{minipage}
\end{center}
\end{figure}

Since we have allowed the 3-tensor indices to have different ranks, we 
can write 
\bea 
&& \la \cO_{ \s_1 , \s_2 , \s_3 } \bar{\cO}_{ \tau_1 , \tau_2 , \tau_3 } \ra
= \sum_{\mu_i \in S_n  }  \sum_{\nu_i \in S_n  }  
 N_1^{ C_{ \s_1 \mu_1 \nu_1 \tau^{-1}_1 } }  N_2^{ C_{ \s_2 \mu_1 \nu_1 \tau^{-1}_2 } }  N_3^{ C_{ \s_3 \mu_1 \nu_1 \tau^{-1}_3} } \cr 
 &&  = \sum_{\mu_i \in S_n  }  \sum_{\nu_i \in S_n  }  
\sum_{\alpha_i  \in S_n  }  
N_1^{ C_{\alpha_1}  }N_2^{ C_{\alpha_2}  } N_3^{ C_{\alpha_3}  } 
\delta (  \s_1 \mu_1  \tau^{-1}_1\nu_1 \alpha_1  ) 
\delta (  \s_2 \mu_2 \tau^{-1}_2 \nu_2 \alpha_2  )  
\delta (  \s_3 \mu_3  \tau^{-1}_3  \nu_3 \alpha_3 )  \cr 
&&   = \sum_{\mu_i \in S_n  }  
\sum_{\nu_i \in S_n  }   
N_1^{ n }N_2^{ n } N_3^{ n   } 
\delta (  \s_1 \mu_1  \tau^{-1}_1 \nu_1 \Omega_1  ) 
\delta(\s_2 \mu_2 \nu_2 \tau^{-1}_2 \Omega_2)
\delta(\s_3 \mu_3 \nu_3 \tau^{-1}_3 \Omega_3) \cr 
&& 
\eea
In the second line, we introduced three extra permutations constrained by delta functions 
to re-write the previous line. 
Note that $ C_{ \alpha } = C_{ \alpha^{-1} } $.   In the third line, 
we have extracted the leading power of $N$ which comes from the permutation 
with the largest number of cycles, namely the identity permutations. 
The $\Omega ( N)$ factor is an element of the group algebra of $S_n$ of the form 
\be
N^{n} \Omega = N^{n} \Big(  1+ \sum_{ \alpha  \in S_n\ses \{1\} } 
\frac{ N^{C_{\alpha}} }{ N^{n} } \alpha \Big).
\ee
This element plays a key role in large $N$ expansions of two dimensional 
YM \cite{GRT1,GRT2}. 
The device of introducing delta functions makes the connection 
to the counting of branched covers transparent.
Thus, we have derived \eqref{corr2pt} stated in Section \ref{sec:correlators}.

\section{GAP and Mathematica codes}
\label{app:gap}

We provide here some programming codes of GAP
and Mathematica. These have allowed
the determination of several sequences in the text. 
The number $Z_d(n)$ of rank $d$ tensor invariants, made with $n$ covariant tensors $T$ and $n$ contravariant tensor $\bar T$, is the one we
primarily focused. Then other numbers are derived from it.   
After entering a given line, the line starting by {\tt (out)} should be obtained. 
  
\bigskip

\noindent{\bf GAP code 1 for $Z_{d}(n)$ and $Z_{d;\,\col}(n)$.}
We provide here a code for evaluating $Z_{3}(n)$, number of
rank 3 tensor invariants, and $Z_{3;\,\col}(n)$ number of 
rank 3 color-symmetrized tensor invariants. 
In the following program, we use the particular value $n=4$.
Changing that parameter $n$ or introducing a procedure
for any finite range of value of $n$ will allow one to  
recover the full sequences \eqref{SumSymMu}, \eqref{r3conX}, \eqref{z3colconX}
and \eqref{r3colorconX} in the text.  Meanwhile changing the rank $d$ of the tensor will require little extra work and allow to find \eqref{ncopdtens} giving, in particular for
$d=4$, \eqref{ran4}. 

The sequence of lines starting by the prompt {\tt gap>} denotes the lines entered. The following lines with {\tt (out)} are the outputs of that entry. The procedure 
starts by the computation of $Z_{3}(n=4)$ using the formula \eqref{gam2}. This allows us to reduce the number of steps because we simply avoid another 
sum over $S_4$. Then, from this, we can evaluate the number of connected invariants $Z^{\rm connect}_{3}(n=4)$ \eqref{r3conX}, the number of colored symmetrized invariants $Z_{3;\,\col}(n=4)$ \eqref{z3colconX} and then the number $Z^{\rm connect}_{3;\,\col}(n=4)$ of color symmetrized connected invariants  \eqref{r3colorconX}. 
Interestingly, in order to obtain connected graphs, we use the
command {\tt IsTransitive (G, [1..4])} checking if the action of the
group {\tt G} on $\{1,2,3,4\}$ is transitive.

{\scriptsize\begin{verbatim}
gap> TT := [ ]; 

(out) [ ]

gap> n := 4;; 
     for tau1 in SymmetricGroup(n) 
           do for tau2 in SymmetricGroup(n) 
               do Add (TT, [tau1, tau2]); 
               od; 
           od;

gap> TT[3];

(out) [ (), (1,2,4) ]

gap> OTT := OrbitsDomain (SymmetricGroup(n), TT, OnPairs);; 
      Ln := Length (OTT);;
	      Print("Z_3(n=4) = ", Ln);

(out) Z_3(n=4) = 43        

gap> OS := [];;  for k in [1..Ln] 
                    do Add (OS , Set(OTT[k]));  
                     od;  

gap> OU := [];; for p in [1..Length(Unique(OS))] 
                    do Add (OU, Unique(OS)[p][1]); 
                     od; 

gap> cnx := [];; for j in [1..Length(Unique(OS))] 
                   do if IsTransitive (Group (OU[j][1], OU[j][2]) , [1..m])   
                         then Add (cnx, OU[j]); 
                       fi; 
                   od;  

gap> Print("Z^{connect}_3(n=4) = " , Length (cnx) );

(out) Z^{connect}_3(n=4) = 26


gap> P23 := function (List2P) 
              local LL ; 
              LL := []; 
              Add (LL , List2P[2]); 
              Add (LL , List2P[1]); 
              return LL; 
        end;

(out) function( List2P ) ... end

gap> P12 := function (List2P) 
              local LL ; 
              LL := []; 
              Add (LL , Inverse ( List2P[1] ) ); 
              Add (LL, Inverse ( List2P[1] ) * List2P[2] ); 
             return LL; 
         end;

(out) function( List2P ) ... end

gap> QROTT := [];; 
        for i in [1 .. Ln] 
        do Add (QROTT, [ ]); 
         od;     
      
       for i in [1.. Ln] 
         do for j in [1 .. Length (OTT[i]) ] 
           do Add (QROTT[i] , OTT[i][j] ); 
		              Add (QROTT[i] , P12 (OTT[i][j])); 
              Add (QROTT[i] , P23 (OTT[i][j])); 
              Add (QROTT[i] , P12( P23 (OTT[i][j]) ) ); 
              Add (QROTT[i] , P23( P12 (OTT[i][j]) ) ); 
              Add (QROTT[i] , P12( P23 (P12 (OTT[i][j]) ))); 
           od; 
         od;


gap> Length (QROTT);

(out) 43

gap> SetQROTT := [];;  for i in [1..Ln] 
                           do Add (SetQROTT, Set (QROTT[i])); 
                             od; 
            LUsq := Length ( Unique (SetQROTT) );;

gap> Print ("Z_{3;color}(n=4)=", LUsq); 

Z_{3;color}(n=4) = 15

gap> UQROTT := [];;  for i in [1..LUsq] 
                        do Add (UQROTT, Unique (SetQROTT)[i][1]);
                         od;

gap> CnX := [];;  for i in [1..LUsq] 
                   do if IsTransitive (Group (UQROTT[i][1], UQROTT[i][2]), [1..n]) 
                       then Add (CnX, UQROTT[i]); 
                      fi; 
                   od;

gap> Print ("Z^{connect}_{3;color}(n=4) = ", Length(CnX));

Z^{connect}_{3;color}(n=4) = 8
\end{verbatim}}

\medskip 

\noindent{\bf Mathematica code 1 for $Z_d(n)$.}
In this paragraph, we provide a Mathematica code for evaluating the number $Z_d(n)$ (denoted {\tt Z[n,d]}) of rank $d$ tensor invariants made
with $2n$ tensors. Specifically, we evaluate $Z_3(n)$ and $Z_4(n)$ for the rank 3 and 4, respectively. We use the built-in function {\tt Count[list, pattern]} which
count the number of element in a {\tt list} matching a {\tt pattern}.  
We also give the code for the generating functions $Z_{d}(x)$
(denoted {\tt Zseries[x,d]}) from which the Plog function $\plog Z_{d}(x)$ (denoted {\tt PLogZ[F,d,x]}) is derived. Then we can obtain the number of connected invariants from the later function using the built-in M\"obius function.

{\scriptsize

\begin{verbatim}
IntegerPartitions [ 4 ] 
IntegerPartitions [ 4 ][[1]]  

(out) {{4}, {3, 1}, {2, 2}, {2, 1, 1}, {1, 1, 1, 1}}
(out) {4}

Count [{1,1}, 2]
Count [{1,1,2}, 1]

(out) 0
(out) 2

Sym [p_ , n_ ] := Product [ i^( Count [p , i] ) ( Count [p , i] )! , {i, 1, n} ]  

Sym [{1, 1} , 2] 

(out) 2

Z [n_ , d_] := Sum [ ( Sym [ IntegerPartitions[n][[i]] , n ] )^(d - 2) , 
                     {i, 1, Length [ IntegerPartitions[n] ] } ]

Zseries [x_, d_] := Sum [ Z [n , d] x^n , {n, 0, 10} ] 

Zseries [x , 3]

(out) 1 + x + 4 x^2 + 11 x^3 + 43 x^4 + 161 x^5 + 901 x^6 + 5579 x^7 + 
 43206 x^8 + 378360 x^9 + 3742738 x^10

Zseries [x , 4]

1 + x + 8 x^2 + 49 x^3 + 681 x^4 + 14721 x^5 + 524137 x^6 + 
 25471105 x^7 + 1628116890 x^8 + 131789656610 x^9 + 13174980291658 x^10


PLog [F_, d_, t_]   := Sum [ MoebiusMu [ k ] / k Log [ F [t^k , d] ] , {k, 1, 10} ]  


Do[ Print[ "Plog[Z(", x, ",", d, ")] = ", Series [ PLog [ Zseries, d, x] , {x , 0, 10} ] ] , {d, 3, 4}]


(out) Plog[Z(x,3)] = x+ 3 x^2 + 7 x^3 + 26 x^4 +97 x^5 + 624 x^6 + 4163 x^7
+ 34470 x^8 + 314493 x^9 + 3202839 x^10+ O[x]^11

(out) Plog[Z(x,4)] = x + 7 x^2 + 41 x^3 + 604 x^4 + 13753 x^5 + 504243 x^6
+ 24824785 x^7 + 1598346352 x^8 + 129958211233 x^9 + 13030565312011 x^10 + O[x]^11

\end{verbatim}
}
\medskip

\noindent{\bf Mathematica code 2 for $S^{(3)}_{[2,1]} $,
$S^{(3)}_{[3]}(n)$, $S^{(4)}_{[3,1]}(n)$ and $S^{(4)}_{[4]}(n)$.} In this paragraph, we provide Mathematica codes useful for the evaluation of $S^{(3)}_{[2,1]} $ and $S^{(3)}_{[3]}(n)$ appearing in
 $Z_{3;\, \col}(n)$ and $S^{(4)}_{[3,1]}(n)$ and $S^{(4)}_{[4]}(n)$ appearing in $Z_{4;\,\col}(n)$. The sums can be programmed  in a very similar way. 

{\scriptsize 
\begin{verbatim}
X = Array [x , 15]

(out) {x[1], x[2], x[3], x[4], x[5], x[6], x[7], x[8], x[9], x[10], x[11], x[12], x[13], x[14], x[15]}

Z [X , t] := Product [ Exp [ t^i  x[i]/i ] , {i, 1, 15} ]

RR =  Table [ x[2 i] -> x[i]^2 ,  {i, 1, 5} ]

(out) {x[2] -> x[1]^2, x[4] -> x[2]^2, x[6] -> x[3]^2, x[8] -> x[4]^2, x[10] -> x[5]^2, x[12] -> x[6]^2, 
x[14] -> x[7]^2}

Z2 [X , t] = Z [X , t] /. RR 

(out)  e^(t x[1] + 1/2 t^2 x[1]^2 + 1/4 t^4 x[2]^2 + 1/3 t^3 x[3] + 1/6 t^6 x[3]^2 + 1/8 t^8 x[4]^2 
+ 1/5 t^5 x[5] + 1/10 t^10 x[5]^2 + 1/12 t^12 x[6]^2 + 1/7 t^7 x[7] + 1/14 t^14 x[7]^2 + 1/9 t^9 x[9] 
+ 1/11 t^11 x[11] + 1/13 t^13 x[13] + 1/15 t^15 x[15])

PP [n_] := IntegerPartitions [n] 

Z2ans [n_] := Coefficient [ Series [ Z2[X , t] , {t, 0, n} ], t^n ]

Z2ans [3] 

(out)  1/3 (2x[1]^3 + x[3])

Symm [q_ , n_] := Product [ i^{ Count[q , i]}(Count[q , i])! , {i, 1, n} ]

Symm [ {2, 2, 1} , 5] 

(out) {8}

CC [n_ , q_] := Coefficient [ Z2ans[n] , Product [x[i]^(Count[q , i]) , {i, 1, n} ]  ]   

CC [ 3 , { 1, 1, 1} ]
CC [ 3 , { 3} ] 
Z2ans [ 3 ] 

(out) 2/3 
(out) 1/3
(out) 1/3 (2 x[1]^3 + x[3])

S2ans [n_] := Sum [ CC [ n , PP[n][[i]] ] * Symm [ PP[n][[i]], n ] , {i, 1, Length [ PP[n] ] } ]

Do [Print ["S_2(", i, ") = ", S2ans[i][[1]]], {i, 10}]  

Table [ S2ans [i] , {i, 1, 10} ]  

(out) {1, 2, 5, 13, 31, 89, 259, 842, 2810, 10020}

\end{verbatim}
}

The Mathematica code for $S^{(3)}_{[3]}(n)$ can be
obtained from the above code by simply replacing 
{\tt RR}  and {\tt S2ans} (by {\tt S3ans}) entries as follows 

{\scriptsize
\begin{verbatim}

RR  =  Table [ x[3 i] -> x[i]^3 , {i, 1, 5} ]

S3ans [n_] := Sum [ ( CC [ n , PP[n][[i]] ] )^2 * Symm [ PP[n][[i]], n ] , {i, 1, Length [ PP [n] ] } ]

Table [ S3ans [i] , {i, 1, 10} ]  

(out) {1, 1, 2, 4, 5, 13, 29, 48, 114, 301}

\end{verbatim}
} 

The Mathematica code for $S^{(4)}_{[3,1]}(n)$ can be
obtained from the above code by substituting 
the definition of {\tt S3ans} as follows 

{\scriptsize
\begin{verbatim}

S3ans [n_] := Sum [ CC [ n , PP[n][[i]] ] * Symm [ PP[n][[i]], n ] , {i, 1, Length [ PP [n] ] } ]

Table [ S3ans [i] , {i, 1, 10} ]  

(out) {1, 2, 4, 12, 27, 103, 391, 1383, 6260, 32704}

\end{verbatim}
}

The Mathematica code for $S^{(4)}_{[4]}(n)$ can be also 
obtained by simply replacing (as well where necessary afterwards)
 {\tt RR}, {\tt Z2} (by {\tt Z4}) and {\tt S2ans} (by {\tt S5ans}) entries as follows 

{\scriptsize
\begin{verbatim}
RR  =  Table [ { x[4 i] -> x[i]^4 , x[4 i - 2 ] -> x[2 i - 1]^2 }, {i, 1, 4} ]

FRR := Flatten [ RR ] 

Z5 [X , t] = Z [X , t] /. FRR 

S5ans [n_] := Sum [ ( CC [ n , PP[n][[i]] ] )^2 * Symm [ PP[n][[i]], n ] , {i, 1, Length [ PP [n] ] } ]

Table [ S5ans [i] , {i, 1, 10} ]  

(out) {1, 2, 3, 11, 27, 93, 233, 978, 3156, 13280} 

\end{verbatim}
}

\noindent{\bf Mathematica code 3 for $S^{(4)}_{[2,1^2]}(n)$.} In this paragraph, we provide Mathematica codes useful for evaluating $S^{(4)}_{[2,1^2]}(n)$ occurring in $Z_{4;\, \col}(n)$.

{\scriptsize
\begin{verbatim}
SymH [n_ , p_] := Product [ (2 j)^( 2 Count [p , 4 j]) Factorial [ 2 Count [p , 4 j] ] , {j, 1, Floor [ n/2 ] } ] 
Product [ (2 j + 1)^( Count [p , 2 j  + 1] + 2 Count [p , 4 j + 2] ) Factorial [Count [ p , 2 j + 1] + 
2 Count [ p , 4 j + 2 ]  ] , {j, 0, Floor [  n/2  ] } ] 

SymH [ 3 , {3} ] 
SymH [ 3 , {1,1,1} ] 
SymH [ 4 , {4} ]

(out) 3
(out) 6
(out) 8

PP [ n_ ] := IntegerPartitions [ n  ] 

Sp2  [ n_ ] := Sum [ SymH [ n , PP[n][[i]] ] , {i , 1, Length [ PP [n] ] } ]

Table [ Sp2 [i] , {i, 1, 10} ]  

(out) {1, 4, 15, 83, 385, 2989, 20559, 203922, 1827640, 21863590}

\end{verbatim}
}

\noindent{\bf Mathematica code 4 for $S^{(4)}_{[2^2]}(n)$.} 
The code for $S^{(4)}_{[2^2]}(n)$ is again very similar to the above code 2
for $S^{(3)}_{[2,1]} $. We simply  remove some lines 
and adjust the final {\tt S2ans} in order to 
evaluate $S^{(4)}_{[2^2]}(n)$. 

{\scriptsize
\begin{verbatim}
X = Array [x , 15] 

Z [X , t] := Product [ Exp [ t^i  x[i]/i ] , {i, 1, 15} ]

RR = Table [ x[2 i] -> x[i]^2, {i, 1, 7} ]

Z2 [X , t] = Z [X , t] /. RR 

PP [ n_ ] := IntegerPartitions [n] 

Z2ans [n_] := Coefficient [ Series [ Z2 [X,  t] , {t, 0, n} ], t^n ]

CC [n_ , q_] := Coefficient [ Z2ans [n] , Product [ x[i]^(Count[q , i]) , {i, 1, n} ] ]  

S4prime [n_] := Sum [ (CC [ n , PP[n][[i]] ])^2 , {i, 1, Length [ PP [n] ] } ]

Table [ S4prime [i] , {i, 1, 10} ]  

(out) {1, 4, 17, 105, 685, 5825, 54013, 585018, 6872522, 90254150}

\end{verbatim}
}

\noindent{\bf Mathematica code 5 for $\cZ_{d; \ncol}(n)$.} 
Here, we provide a program which yields \eqref{noncol2} 
and \eqref{noncol3}.  

{\scriptsize
\begin{verbatim}
PP [n_] := IntegerPartitions  [ n ] 

CC [d_ , n_] := Sum [ Product [ i^( (d - 2) Count [PP[n][[j]] , i] ) 
* (d Count [PP[n][[j]], i ])! / (Count [PP[n][[j]] , i ]!)^2 , {i, 1, n}] , {j, 1, Length [ PP[n] ] } ] 

Table [ CC [2, j] , {j, 1, 10} ] 

(out) {2, 8, 26, 94, 326, 1196, 4358, 16248, 60854, 230184}

Table [ CC [3, j] , {j, 1, 10}]

(out) {6, 192, 10170, 834612, 90939630, 12360636540, 2012440468938, 381799921738584}

\end{verbatim}
}

\noindent{\bf Mathematica code 6 for $\cZ_{d; \sym}(n)$.} 
The following codes allow us to obtains the sequences \eqref{d213}
\eqref{d38} and \eqref{d47} for $\cZ_{d; \sym}(n)$, for any
rank $d\geq 2$ and order $n\geq 1$. 

{\scriptsize
\begin{verbatim}
X = Array [x , 15]

PP [n_] := IntegerPartitions [n] 

Sym [q_  , n_] := Product [ i^(Count [q , i]) Count [q , i] ! , {i , 1, n}]

Symd [X, k_, q_, d_] := Product [ ( X[[k *l ]] / l )^(Count [q , l])/( Count [ q , l ] !) , {l, 1, d} ]

Z  [X, t, d_] := Product [ Exp [ ( t^i /i ) * Sum[ Symd [X, i, PP [d][[j]], d], {j, 1, Length [ PP [d] ]}] ] ,
                  {i, 1, 15} ]

Zprim [ n_, d_ ] := Coefficient [ Series [ Z [X,  t, d] , {t, 0, n} ] , t^n ]

CC[ n_ , q_ , d_] := Coefficient [ Zprim [n , d] , Product [ X[[i]]^( Count [q , i] ) , {i, 1, dn} ] ] 

Zdsym [ n_, d_ ]  := Sum [ ( CC [n , PP [d n][[i]], d])^2 * Sym [ PP [n d][[i]] , dn ] ,
           {i, 1, Length [ PP [nd] ] } ]

Table [ Zdsym [i, 2] , {i, 1, 13} ]

(out) {1, 2, 3, 5, 7, 11, 15, 22, 30, 42, 56, 77, 101}

Table [ Zdsym [i, 3] , {i, 1, 7} ]

(out) {1, 2, 5, 12, 31, 103, 383, 1731}
\end{verbatim}
}

\end{document}